\documentclass[11pt,dvips,twoside,letterpaper]{article}
\usepackage{pslatex}
\usepackage{fancyhdr}
\usepackage{graphicx}
\usepackage{geometry}
\usepackage{amsmath}
\usepackage{amssymb}
\usepackage{amsfonts}
\usepackage{amsthm,amscd}

\def\figurename{Figure}
\makeatletter
\renewcommand{\fnum@figure}[1]{\figurename~\thefigure.}
\makeatother
\def\tablename{Table}
\makeatletter
\renewcommand{\fnum@table}[1]{\tablename~\thetable.}
\makeatother
\newtheorem{theorem}{Theorem}[section]

\theoremstyle{definition}

\theoremstyle{remark}
\newtheorem{remark}[theorem]{Remark}
\numberwithin{equation}{section}

\input{tcilatex}

\begin{document}

\title{\bfseries%
\scshape{On the spherically symmetric Einstein-Yang-Mills-Higgs
equations in Bondi coordinates}}
\author{\bfseries\scshape Calvin Tadmon$^{1,2}$\thanks{%
E-mail address: tadmonc@yahoo.fr; calvin.tadmon@up.ac.za} \\
%EndAName
$^{1}$Department of Mathematics and Computer Science, \\
University of Dschang, P.\ O.\ Box 67, Dschang, Cameroon\\
$^{2}$Department of Mathematics and Applied Mathematics,\\
University of Pretoria, Pretoria 0002, South Africa\\
\bfseries\scshape Sophonie Blaise Tchapnda$^{3}$\thanks{%
E-mail address: sophonieblaise@yahoo.com}\\
$^{3}$Department of Mathematics, University of Yaounde 1, \\
P.\ O.\ Box 812, Yaounde, Cameroon}
\date{}
\maketitle

\begin{abstract}
We revisit and generalize, to the Einstein-Yang-Mills-Higgs system, previous
results of \ D. Christodoulou and D. Chae concerning global solutions for
the Einstein-scalar field and the Einstein-Maxwell-Higgs equations. The
novelty of the present work is twofold. For one thing the assumption on the
self-interaction potential is improved. For another thing explanation is
furnished why the solutions obtained here and those proved by Chae for the
Einstein-Maxwell-Higgs decay more slowly than those established by
Christodoulou in the case of self-gravitating scalar fields. Actually this
latter phenomenon stems from the non-vanishing local charge in
Einstein-Maxwell-Higgs and Einstein-Yang-Mills-Higgs models.
\end{abstract}

%\vskip 0.4in

% ------- [First Page Running Head] - place it immediately after title! ------
\thispagestyle{empty} \fancyhead{} \fancyfoot{} \renewcommand{%
\headrulewidth}{0pt}


\vspace{0.08in} \noindent \textbf{Keywords}: Global solution,
Einstein-Yang-Mills-Higgs equations, spherically symmetric, Bondi
coordinates.

\section{Introduction}

The Yang-Mills-Higgs (YMH) field equations arise in elementary particle
physics. The Yang-Mills equations appear as the generalization of the
classical Maxwell equations where ordinary derivatives are replaced by
covariant derivatives. The YMH equations are nonlinear partial differential
equations (PDE) that are conformally invariant and gauge invariant. This
latter property has been exploited by Eardley and Moncrief \cite{12, 13} to
make a remarkable contribution to the area of PDE in proving global
existence for these equations in $4-$dimensional Minkowski space.

When the YMH equations are coupled to gravity, such a global existence
result is yet to be proven without any symmetry assumption. Since this
problem is a very difficult one, it makes sense to start by investigating it
with some symmetry assumption. In this paper we prove a global existence
result for the Eintein-Yang-Mills-Higgs (EYMH) equations under the
assumption of spherical symmetry.

The issue of proving global existence results for gravitational and matter
field equations is of interest in mathematical relativity and in the area of
(PDE). In general relativity, the global existence problem is important
since it is a reformulation of the cosmic censorship conjecture rendering
this more amenable to direct analytic attack (see \cite{14}).

Much work has been done in the past concerning the EYMH equations. Let us
mention for instance that the spontaneous compactification of space in EYMH
model has been studied in \cite{9}. A canonical formulation of the
spherically symmetric EYMH system for a general gauge group has been found
in \cite{8}. Regular localized solutions of the electric or magnetic type
were found in \cite{7}. Stimulated by the discovery of globally regular
solutions of the EYM equations, numerically by Bartnik and McKinnon \cite{1}%
, many authors published a considerable number of papers dealing with the
static spherically or axially symmetric EYMH equations. Most of these papers
concern asymptotically flat, smooth particle-like and black hole solutions
(see e.g. \cite{16}, \cite{19}, \cite{20}, \cite{22}, \cite{27}) although
some cosmological solutions have been found for instance in \cite{2, 15}.

As far as rigorous existence proofs are concerned, K\"{u}nzle and Oliynyk 
\cite{21} used differential geometry techniques to investigate the static
spherically symmetric EYMH system. Local existence to the characteristic
Cauchy problem for EYMH without any symmetry assumption has been proved
recently by Dossa and Tadmon \cite{10, 11}. More recently, Vacaru \cite{26}
has written the Einstein field equations in variables adapted to
nonholonomic $2+2$ splitting and has applied this geometric technique for
constructing off-diagonal exact solutions of EYMH equations.

In the present paper, as mentioned\ above, we investigate the initial value
problem for the spherically symmetric EYMH equations. We write the system of
equations in the so-called Bondi coordinates so as to obtain a
characteristic Cauchy problem, then we prove global existence and some decay
property of solutions. This setting has been used before by Christodoulou 
\cite{6} for the Einstein-scalar field system and by Chae \cite{3, 4}
respectively for Einstein-Klein-Gordon and Einstein-Maxwell-Higgs equations.
To reach our goal, we take advantage of the tools set up in \cite{6} to
reduce the problem to that of the resolution of a nonlinear evolution system
of two PDE. By comparison with the works in \cite{4, 6} where a single
evolution equation were dealt with, supplementary difficulties arise: here
the number of terms to estimate is much more higher and some additional
terms have to be handled meticulously. By dint of arduous calculations
combined with several handy mathematical tools, we implement a fixed point
method to arrive, under a more general assumption on the self-interaction
potential, at a global existence and uniqueness result for the spherically
symmetric EYMH system with appropriate initial data. Furthermore we show
that this solution decays more slowly than the one obtained in \cite{6}. A
thorough examination reveals that this latter phenomenon stems from the
presence of the local charge $Q$. Some questions raised by Chae \cite{4} are
thus answered. It could be interesting to see whether the decay properties
of the solution mentioned above can be explained by using compactification
techniques due to Penrose \cite{23}. In final, we obtained a generalization
and improvement of the results in \cite{4, 6}. The main result of the
present paper encompasses the EYMH system with zero self-interaction
potential and the Einstein-Klein-Gordon (EKG) system as well. It is worth
noting that this paper is the corrected and detailed version of the Note 
\cite{24}.

The present work is organized as follows. Section $2$ is devoted to the
derivation of EYMH equations under the spherical symmetry assumption. In
section $3$ we show that the equations reduce to a nonlinear evolution
system. In the last section we state and prove the main result of the
present investigation.

\section{The Einstein-Yang-Mills-Higgs system}

\subsection{Equations of motion}

% ------------ [Running Heads - for odd and even pages] - please insert it only on page 2!
\pagestyle{fancy} \fancyhead{} \fancyhead[EC]{C. Tadmon, S. B. Tchapnda} %
\fancyhead[EL,OR]{\thepage} 
\fancyhead[OC]{Global solution of the
spherically symmetric EYMH equations} \fancyfoot{} \renewcommand%
\headrulewidth{0.5pt}

Throughout the paper, unless otherwise is stated, Einstein convention is
used, e.g., $w_{c}^{d}h^{c}=\underset{c}{\sum }w_{c}^{d}h^{c}$. We
concentrate on the $\mathfrak{su}(2)-$EYMH and assume that the Yang-Mills
field is in the adjoint representation of $\mathfrak{su}(2)$ while the Higgs
fields is in the fundamental representation of $\mathfrak{su}(2)$. The basic
elements of the $\mathfrak{su}(2)-$EYMH system consist of a quadruplet $%
\left( \mathcal{M},g,A,\Phi \right) $, where $\mathcal{M}$ is a $4D$
space-time manifold equipped with a metric $g$; $A$ is a $1-$form, called
the Yang-Mills potential, defined on $\mathcal{M}$ with values in the Lie
algebra $\mathfrak{su}(2)$ of the Lie group $SU(2)$; $\Phi $ is a
scalar-multiplet field, called the Higgs field, defined on $\mathcal{M}$.
The intrinsic fields equations for the $\mathfrak{su}(2)-$EYMH model are
obtained from an intrinsic Lagrangian. For the EYMH model with a Yang-Mills
field in the adjoint representation of $\mathfrak{su}(2)$ and a complex
Higgs field in the fundamental representation of $\mathfrak{su}(2)$ the $4D$
action is%
\begin{equation}
S_{EYMH}=\int \left( L_{E}+L_{YM}+L_{H}\right) \left[ -\det (g_{\alpha \beta
})\right] ^{\frac{1}{2}}d^{4}x,  \tag{2.1}  \label{2.1}
\end{equation}%
where $L_{E}$, $L_{YM}$, and $L_{H}$ denote, respectively, the gravity part,
the Yang-Mills field part, and the Higgs field part of the Lagrangian
density of the EYMH model. To avoid cumbersome coefficients in the final
form of the fields equations we write the above Lagrangian densities as
follows%
\begin{equation*}
\begin{array}{l}
L_{E}=-\frac{g^{\alpha \beta }R_{\alpha \beta }}{16\pi G}, \\ 
L_{YM}=\frac{1}{32\pi G}g^{\gamma \beta }g^{\rho \sigma }F_{\gamma \rho
}.F_{\beta \sigma }, \\ 
L_{H}=\frac{1}{8\pi G}\left[ g^{\gamma \beta }(\widehat{\nabla }_{\gamma
}\Phi )^{\dag }\widehat{\nabla }_{\beta }\Phi +V\left( \Phi ^{\dag }\Phi
\right) \right] ,%
\end{array}%
\end{equation*}%
where $\left( R_{\alpha \beta }\right) $ is the Ricci curvature relative to
the space-time metric, $G$ is the universal Newtonian gravitational
constant. $\left( F_{\alpha \beta }\right) $ represents the Yang-Mills
strength field $F$, which\ is a $\mathfrak{su}(2)$-valued antisymmetric $2-$%
form of type $Ad,$ defined on $\mathcal{M}$. $F$ is related to the unknown\
Yang-Mills potential $A$ as follows%
\begin{equation}
F_{\alpha \beta }^{I}=\nabla _{\alpha }A_{\beta }^{I}-\nabla _{\beta
}A_{\alpha }^{I}+\left[ A_{\alpha },A_{\beta }\right] ^{I},  \tag{2.2}
\label{2.2}
\end{equation}%
with $\left[ A_{\alpha },A_{\beta }\right] ^{I}=\varepsilon
_{JK}^{I}A_{\alpha }^{J}A_{\beta }^{K}$, where $F_{\alpha \beta }^{I}$ and $%
A_{\alpha }^{I}$ are the respective components of $F$ and $A$ in local
coordinates $\left( x^{\alpha }\right) _{\alpha =0,...,3}$ on $\mathcal{M}$
and basis $\left( T_{I}\right) _{I=1,...,3}$ of $\mathfrak{su}(2)$, i.e.,%
\begin{equation*}
F=\frac{1}{2}\left( F_{\alpha \beta }^{I}T_{I}\right) dx^{\alpha }\wedge
dx^{\beta },\quad A=\left( A_{\alpha }^{I}T_{I}\right) dx^{\alpha }.
\end{equation*}%
$\nabla $ denotes the covariant derivative w.r.t. the space-time metric $g$,$%
\ \left[ ,\right] $ denote the Lie brackets of the Lie algebra $\mathfrak{su}%
(2)$, and$\ \varepsilon _{JK}^{I}$ are the structure constants of $\mathfrak{%
su}(2)$. We work within the basis $\left( T_{I}\right) _{I=1,...,3}$ of $%
\mathfrak{su}(2)$, with 
\begin{equation*}
T_{1}=i\frac{\sigma _{1}}{2},\quad T_{2}=-i\frac{\sigma _{2}}{2},\quad
T_{3}=i\frac{\sigma _{3}}{2},
\end{equation*}%
where $\left( \sigma _{I}\right) _{I=1,2,3}$ are the conventional Pauli spin
matrices defined by%
\begin{equation*}
\sigma _{1}=\left( 
\begin{array}{cc}
0 & 1 \\ 
1 & 0%
\end{array}%
\right) ,\quad \sigma _{2}=\left( 
\begin{array}{cc}
0 & -i \\ 
i & 0%
\end{array}%
\right) ,\quad \sigma _{3}=\left( 
\begin{array}{cc}
1 & 0 \\ 
0 & -1%
\end{array}%
\right) .
\end{equation*}%
It is obvious to see that in the basis $\left( T_{I}\right) _{I=1,...,3}$
defined above, the structure constants are given by 
\begin{equation*}
\varepsilon _{JK}^{I}=\left\{ 
\begin{array}{c}
1\text{ if }IJK\in \left\{ 123,231,312\right\} , \\ 
-1\text{ if }IJK\in \left\{ 132,213,321\right\} , \\ 
0\text{ if }I=J\text{ or }I=K\text{ or }J=K.%
\end{array}%
\right. 
\end{equation*}%
The dot $^{\text{\textquotedblleft }}.^{\text{\textquotedblright }}$ denotes
the $Ad$-invariant non degenerate scalar product of $\mathfrak{su}(2)$
defined by%
\begin{equation*}
f.k=\overset{3}{\underset{I=1}{\sum }}f^{I}k^{I},\quad f\in \mathfrak{su}%
(2),\quad k\in \mathfrak{su}(2).
\end{equation*}%
It is worth noting that the $Ad$-invariant non degenerate scalar product $%
^{``}.^{\textquotedblright }$ enjoys the following property%
\begin{equation}
f.\left[ k,l\right] =\left[ f,k\right] .l,\text{ }\forall f,k,l\in \mathfrak{%
su}(2).  \tag{2.3}  \label{2.3}
\end{equation}%
$\widehat{\nabla }_{\alpha }\Phi $ is the gauge covariant derivative of the
complex doublet Higgs field, defined by%
\begin{equation}
\widehat{\nabla }_{\alpha }\Phi =\frac{\partial \Phi }{\partial x^{\alpha }}%
-iA_{\alpha }^{I}\frac{\sigma _{I}}{2}\Phi .  \tag{2.4}  \label{2.4}
\end{equation}%
We use $\Phi ^{\dag }$ for the hermitian conjugate of $\Phi $, i.e., $\Phi
^{\dag }$ is the transpose of $\Phi ^{\ast }$, where $\Phi ^{\ast }$ is the
complex conjugate of $\Phi $. $V$ is a real function defined on $\left[
0,\infty \right) $, often called the self-interaction potential, with
derivative $V^{\prime }$.

We are now in the position to derive the EYMH equations. Variation of the
action $\left( \ref{2.1}\right) $ w.r.t. $g^{\alpha \beta }$, $A_{\alpha }$
and $\Phi ^{\dag }$ yields the following equations of motion (\ see \cite{5,
10, 11, 24})%
\begin{equation}
\begin{array}{l}
R_{\alpha \beta }-\frac{1}{2}g_{\alpha \beta }R=T_{\alpha \beta }, \\ 
g^{\lambda \mu }\left( \nabla _{\lambda }F_{\mu \alpha }+\left[ A_{\lambda
},F_{\mu \alpha }\right] \right) =J_{\alpha }, \\ 
g^{\lambda \mu }\widehat{\nabla }_{\lambda }\widehat{\nabla }_{\mu }\Phi
=V^{\prime }\left( \Phi ^{\dag }\Phi \right) \Phi ,%
\end{array}
\tag{2.5}  \label{2.5}
\end{equation}%
where $T_{\alpha \beta }$ are\ the\ components\ of the
energy-momentum-stress tensor, given by%
\begin{equation}
\begin{array}{l}
T_{\alpha \beta }=g^{\rho \sigma }F_{\alpha \rho }.F_{\beta \sigma }-\frac{1%
}{4}g_{\alpha \beta }g^{\sigma \lambda }g^{\rho \mu }F_{\sigma \rho
}.F_{\lambda \mu } \\ 
\text{ \ \ \ \ \ \ \ \ \ \ }+(\widehat{\nabla }_{\alpha }\Phi )^{\dag }%
\widehat{\nabla }_{\beta }\Phi +(\widehat{\nabla }_{\beta }\Phi )^{\dag }%
\widehat{\nabla }_{\alpha }\Phi -g_{\alpha \beta }\left( g^{\sigma \rho }(%
\widehat{\nabla }_{\sigma }\Phi )^{\dag }\widehat{\nabla }_{\rho }\Phi
+V\left( \Phi ^{\dag }\Phi \right) \right) .%
\end{array}
\tag{2.6}  \label{2.6}
\end{equation}%
$J_{\alpha }$ are the components of the Yang-Mills current, given by 
\begin{equation}
J_{\alpha }^{I}=\Phi ^{\dag }S^{I}\widehat{\nabla }_{\alpha }\Phi -(\widehat{%
\nabla }_{\alpha }\Phi )^{\dag }S^{I}\Phi ,\quad I=1,2,3,\quad \alpha
=0,1,2,3;  \tag{2.7}  \label{2.7}
\end{equation}%
where 
\begin{equation*}
S^{I}=i\frac{\sigma _{I}}{2}.
\end{equation*}

\subsection{The spherically symmetric ans\"{a}tze and fields equations}

We will work in a Bondi coordinates system $\left( x^{\alpha }\right)
=\left( u,r,\theta ,\varphi \right) $ on $\mathbb{R}^{4}$, used in series of
works by D.\ Christodoulou \cite{6} and D.\ Chae \cite{3, 4}, where $u$ is a
retarded time coordinate and $r$ is a radial coordinate. In this coordinate
system the most general form for the spherically symmetric metric could be
written as follows%
\begin{equation}
ds^{2}=-e^{2\nu }du^{2}-2e^{\nu +\lambda }dudr+r^{2}\left( d\theta ^{2}+\sin
^{2}\theta d\varphi ^{2}\right) ,  \tag{2.8}  \label{2.8}
\end{equation}%
where $\nu $ and $\lambda $ are real\ functions of $u$ and $r$ only.

The general form for the spherically symmetric Yang-Mills potential in the
adjoint representation of $\mathfrak{su}(2)$\ could be written as follows
(see \cite{5, 10, 11})%
\begin{equation}
A=aT_{3}du,  \tag{2.9}  \label{2.9}
\end{equation}%
i.e., $A_{0}=aT_{3}$, $A_{1}=A_{2}=A_{3}=0$, where $a$ is a function of $u$
and $r$ only.

We will use the following ansatz for the spherically symmetric Higgs field
in the fundamental representation of $\mathfrak{su}(2)$%
\begin{equation}
\Phi =\left( 
\begin{array}{c}
0 \\ 
\psi +i\xi%
\end{array}%
\right) ,  \tag{2.10}  \label{2.10}
\end{equation}%
where $\psi $ and $\xi $ are real functions of $u$ and $r$ only.

After some tedious and lengthy calculation we find out from the ansatz $%
\left( \ref{2.8}\right) $ that, in the coordinates system $\left( x^{\alpha
}\right) _{\alpha =0,...,3}=\left( u,r,\theta ,\varphi \right) $, the
non-vanishing Christoffel symbols are%
\begin{equation}
\begin{array}{l}
\Gamma _{00}^{0}=\overset{\cdot }{\lambda }+\overset{\cdot }{\nu }-\nu
^{\prime }e^{\nu -\lambda },\quad \Gamma _{22}^{0}=re^{-\nu -\lambda },\quad
\Gamma _{33}^{0}=\Gamma _{22}^{0}\sin ^{2}\theta , \\ 
\Gamma _{00}^{1}=-\overset{\cdot }{\lambda }e^{\nu -\lambda }+\nu ^{\prime
}e^{2\left( \nu -\lambda \right) }\quad ,\quad \Gamma _{01}^{1}=\nu ^{\prime
}e^{\nu -\lambda }, \\ 
\Gamma _{11}^{1}=\lambda ^{\prime }+\nu ^{\prime },\quad \Gamma
_{22}^{1}=-re^{-2\lambda },\quad \Gamma _{33}^{1}=\Gamma _{22}^{1}\sin
^{2}\theta , \\ 
\Gamma _{12}^{2}=r^{-1},\quad \Gamma _{33}^{2}=-\sin \theta \cos \theta
,\quad \Gamma _{13}^{3}=r^{-1},\quad \Gamma _{23}^{3}=\sin ^{-1}\theta \cos
\theta .%
\end{array}
\tag{2.11}  \label{2.11}
\end{equation}%
Here and throughout the paper the dot $\overset{\cdot }{}$ denotes
differentiation with respect to $u$ while the prime $^{\prime }$\ means
differentiation with respect to $r$. E.g., $\overset{\cdot }{\lambda }=\frac{%
\partial \lambda }{\partial u}$, $\nu ^{\prime }=\frac{\partial \nu }{%
\partial r}$.

The relevant components of the Ricci tensor, calculated\ from\ $\left( \ref%
{2.8}\right) $ and $\left( \ref{2.11}\right) $, are\ found\ to\ be%
\begin{equation}
\begin{array}{l}
R_{00}=e^{2\left( \nu -\lambda \right) }\left[ v^{\prime \prime }+\nu
^{\prime }\left( \nu ^{\prime }-\lambda ^{\prime }+2r^{-1}\right) \right]
-e^{\nu -\lambda }\left[ \left( \overset{\cdot }{\lambda }\right) ^{\prime
}+\left( \overset{\cdot }{\nu }\right) ^{\prime }+2r^{-1}\overset{\cdot }{%
\lambda }\right] , \\ 
R_{01}=e^{\nu -\lambda }\left[ v^{\prime \prime }+\nu ^{\prime }\left( \nu
^{\prime }-\lambda ^{\prime }+2r^{-1}\right) \right] -\left[ \left( \overset{%
\cdot }{\lambda }\right) ^{\prime }+\left( \overset{\cdot }{\nu }\right)
^{\prime }\right] ,\quad R_{11}=2r^{-1}\left( \lambda ^{\prime }+\nu
^{\prime }\right) , \\ 
R_{22}=1-e^{-2\lambda }\left[ 1+r\left( \nu ^{\prime }-\lambda ^{\prime
}\right) \right] ,\quad R_{33}=R_{22}\sin ^{2}\theta .%
\end{array}
\tag{2.12}  \label{2.12}
\end{equation}%
From $\left( \ref{2.8}\right) $ and $\left( \ref{2.12}\right) $ we deduce
that the scalar curvature is given by%
\begin{equation}
R=2e^{-\lambda -\nu }\left[ \left( \overset{\cdot }{\lambda }\right)
^{\prime }+\left( \overset{\cdot }{\nu }\right) ^{\prime }\right]
-2e^{-2\lambda }\left[ v^{\prime \prime }+\left( \nu ^{\prime }-\lambda
^{\prime }\right) \left( \nu ^{\prime }+2r^{-1}\right) +r^{-2}\right]
+2r^{-2}.  \tag{2.13}  \label{2.13}
\end{equation}%
From $\left( \ref{2.2}\right) $ and the ansatz $\left( \ref{2.9}\right) $ it
follows that the relevant component of the Yang-Mills strength field is%
\begin{equation}
F_{01}=-a^{\prime }T_{3}.  \tag{2.14}  \label{2.14}
\end{equation}%
The non-vanishing components of the Yang-Mills current, calculated\ from\ $%
\left( \ref{2.4}\right) $ and $\left( \ref{2.7}\right) $, are found\ to\ be%
\begin{equation}
J_{0}=\left[ \psi \overset{\cdot }{\xi }-\xi \overset{\cdot }{\psi }+\frac{1%
}{2}a\left( \psi ^{2}+\xi ^{2}\right) \right] T_{3},\quad J_{1}=\left( \psi
\xi ^{\prime }-\xi \psi ^{\prime }\right) T_{3}.  \tag{2.15}  \label{2.15}
\end{equation}%
Lengthy\ calculation, using $\left( \ref{2.4}\right) $, $\left( \ref{2.8}%
\right) $ and $\left( \ref{2.14}\right) $ in $\left( \ref{2.6}\right) $,
yields\ the following\ non-vanishing components of the
energy-momentum-stress tensor 
\begin{equation}
\begin{array}{l}
T_{00}=\frac{1}{2}e^{-2\lambda }\left( a^{\prime }\right) ^{2}+2\left[
\left( \overset{\cdot }{\psi }-\frac{1}{2}a\xi \right) ^{2}+\left( \overset{%
\cdot }{\xi }+\frac{1}{2}a\psi \right) ^{2}\right] -2e^{\nu -\lambda }\left[
\left( \overset{\cdot }{\psi }-\frac{1}{2}a\xi \right) \psi ^{\prime
}+\left( \overset{\cdot }{\xi }+\frac{1}{2}a\psi \right) \xi ^{\prime }%
\right] \\ 
\text{ \ \ \ \ \ \ \ \ \ }+e^{2\nu }V+e^{2\left( \nu -\lambda \right) }\left[
\left( \psi ^{\prime }\right) ^{2}+\left( \xi ^{\prime }\right) ^{2}\right] ,
\\ 
T_{01}=\frac{1}{2}e^{-\lambda -\nu }\left( a^{\prime }\right)
^{2}+e^{\lambda +\nu }V+e^{\nu -\lambda }\left[ \left( \psi ^{\prime
}\right) ^{2}+\left( \xi ^{\prime }\right) ^{2}\right] , \\ 
T_{11}=2\left[ \left( \psi ^{\prime }\right) ^{2}+\left( \xi ^{\prime
}\right) ^{2}\right] , \\ 
T_{22}=r^{2}\left\{ \frac{1}{2}e^{-2\left( \lambda +\nu \right) }\left(
a^{\prime }\right) ^{2}+2e^{-\lambda -\nu }\left[ \left( \overset{\cdot }{%
\psi }-\frac{1}{2}a\xi \right) \psi ^{\prime }+\left( \overset{\cdot }{\xi }+%
\frac{1}{2}a\psi \right) \xi ^{\prime }\right] -e^{-2\lambda }\left[ \left(
\psi ^{\prime }\right) ^{2}+\left( \xi ^{\prime }\right) ^{2}\right]
-V\right\} , \\ 
T_{33}=T_{22}\sin ^{2}\theta .%
\end{array}
\tag{2.16}  \label{2.16}
\end{equation}%
Writing roughly the EYMH fields equations in the coordinates system $\left(
u,r,\theta ,\varphi \right) $ yields a system of PDE which is very difficult
to handle. Ahead of overcoming this toughness we introduce, as in \cite{6},
a null tetrad $\left( e_{\alpha }\right) _{\alpha =0,...,3}$ defined by%
\begin{equation}
\begin{array}{l}
e_{0}=e^{-\nu }\frac{\partial }{\partial u}-\frac{1}{2}e^{-\lambda }\frac{%
\partial }{\partial r},\quad e_{1}=e^{-\lambda }\frac{\partial }{\partial r},
\\ 
e_{2}=c\frac{\partial }{\partial \theta }+d\frac{\partial }{\partial \varphi 
},\quad e_{3}=h\frac{\partial }{\partial \theta }+k\frac{\partial }{\partial
\varphi },%
\end{array}
\tag{2.17}  \label{2.17}
\end{equation}%
where the functions $c,$ $d,$ $h$ and $k$ are such that $\left(
e_{2},e_{3}\right) $ is a locally defined orthonormal frame on the unit $2-$%
sphere; this implies that%
\begin{equation}
c^{2}+d^{2}\sin ^{2}\theta =r^{-2},\quad h^{2}+k^{2}\sin ^{2}\theta
=r^{-2},\quad ch+dk\sin ^{2}\theta =0.  \tag{2.18}  \label{2.18}
\end{equation}%
In the null tetrad $\left( e_{\alpha }\right) _{\alpha =0,...,3}$ the
relevant components of the space-time metric are computed from\ $\left( \ref%
{2.8}\right) $, $\left( \ref{2.17}\right) $ and $\left( \ref{2.18}\right) $.
These are%
\begin{equation}
g\left( e_{0},e_{1}\right) =-1,\quad g\left( e_{2},e_{2}\right) =g\left(
e_{3},e_{3}\right) =1.  \tag{2.19}  \label{2.19}
\end{equation}%
In the null tetrad $\left( e_{\alpha }\right) _{\alpha =0,...,3}$ the
following\ relevant components of the Ricci tensor are calculated from\ $%
\left( \ref{2.12}\right) $, $\left( \ref{2.17}\right) $ and $\left( \ref%
{2.18}\right) $.%
\begin{equation}
\begin{array}{l}
R\left( e_{0},e_{0}\right) =r^{-1}\left[ \frac{1}{2}e^{-2\lambda }\left(
\lambda ^{\prime }+\nu ^{\prime }\right) -2e^{-\lambda -\nu }\overset{\cdot }%
{\lambda }\right] , \\ 
R\left( e_{0},e_{1}\right) =e^{-2\lambda }\left[ v^{\prime \prime }+\left(
\nu ^{\prime }-\lambda ^{\prime }\right) \left( \nu ^{\prime }+r^{-1}\right) %
\right] -e^{-\lambda -\nu }\left[ \left( \overset{\cdot }{\lambda }\right)
^{\prime }+\left( \overset{\cdot }{\nu }\right) ^{\prime }\right] , \\ 
R\left( e_{1},e_{1}\right) =2r^{-1}e^{-2\lambda }\left( \lambda ^{\prime
}+\nu ^{\prime }\right) , \\ 
R\left( e_{2},e_{2}\right) =r^{-2}\left\{ 1-e^{-2\lambda }\left[ 1+r\left(
\nu ^{\prime }-\lambda ^{\prime }\right) \right] \right\} , \\ 
R\left( e_{3},e_{3}\right) =R\left( e_{2},e_{2}\right) .%
\end{array}
\tag{2.20}  \label{2.20}
\end{equation}%
In the null tetrad $\left( e_{\alpha }\right) _{\alpha =0,...,3}$ the
following\ relevant components of the energy-momentum-stress tensor are
calculated from\ $\left( \ref{2.16}\right) $, $\left( \ref{2.17}\right) $
and $\left( \ref{2.18}\right) $.%
\begin{equation}
\begin{array}{l}
T\left( e_{0},e_{0}\right) =2e^{-2\nu }\left[ \left( \overset{\cdot }{\psi }-%
\frac{1}{2}a\xi \right) ^{2}+\left( \overset{\cdot }{\xi }+\frac{1}{2}a\psi
\right) ^{2}\right] -2e^{-\lambda -\nu }\left[ \left( \overset{\cdot }{\psi }%
-\frac{1}{2}a\xi \right) \psi ^{\prime }+\left( \overset{\cdot }{\xi }+\frac{%
1}{2}a\psi \right) \xi ^{\prime }\right] \\ 
\text{ \ \ \ \ \ \ \ \ \ \ \ \ \ \ \ \ \ }+\frac{1}{2}e^{-2\lambda }\left[
\left( \psi ^{\prime }\right) ^{2}+\left( \xi ^{\prime }\right) ^{2}\right] ,
\\ 
T\left( e_{0},e_{1}\right) =\frac{1}{2}e^{-2\left( \lambda +\nu \right)
}\left( a^{\prime }\right) ^{2}+V, \\ 
T\left( e_{1},e_{1}\right) =2e^{-2\lambda }\left[ \left( \psi ^{\prime
}\right) ^{2}+\left( \xi ^{\prime }\right) ^{2}\right] , \\ 
T\left( e_{2},e_{2}\right) =\frac{1}{2}e^{-2\left( \lambda +\nu \right)
}\left( a^{\prime }\right) ^{2}+2e^{-\lambda -\nu }\left[ \left( \overset{%
\cdot }{\psi }-\frac{1}{2}a\xi \right) \psi ^{\prime }+\left( \overset{\cdot 
}{\xi }+\frac{1}{2}a\psi \right) \xi ^{\prime }\right] -e^{-2\lambda }\left[
\left( \psi ^{\prime }\right) ^{2}+\left( \xi ^{\prime }\right) ^{2}\right]
-V, \\ 
T\left( e_{3},e_{3}\right) =T\left( e_{2},e_{2}\right) .%
\end{array}
\tag{2.21}  \label{2.21}
\end{equation}%
Now, using\ $\left( \ref{2.19}\right) $, $\left( \ref{2.20}\right) $ and $%
\left( \ref{2.21}\right) $ in\ $\left( \ref{2.5}\right) $, the Einstein
fields equations 
\begin{equation}
R\left( e_{\alpha },e_{\beta }\right) -\frac{1}{2}Rg\left( e_{\alpha
},e_{\beta }\right) =T\left( e_{\alpha },e_{\beta }\right)  \tag{2.22}
\label{2.22}
\end{equation}%
in the null tetrad $\left( e_{\alpha }\right) _{\alpha =0,...,3}$ are
equivalent to the following equations%
\begin{equation}
\begin{array}{l}
\frac{1}{2}\left( \lambda ^{\prime }+\nu ^{\prime }\right) -2e^{\lambda -\nu
}\overset{\cdot }{\lambda }\text{ }=2re^{2\left( \lambda -\nu \right) }\left[
\left( \overset{\cdot }{\psi }-\frac{1}{2}a\xi \right) ^{2}+\left( \overset{%
\cdot }{\xi }+\frac{1}{2}a\psi \right) ^{2}\right] \\ 
\text{ \ \ \ \ \ \ \ \ \ \ \ \ \ \ \ \ \ \ \ \ \ \ \ \ \ \ \ \ \ \ \ \ \ \ \ 
}-2re^{\lambda -\nu }\left[ \left( \overset{\cdot }{\psi }-\frac{1}{2}a\xi
\right) \psi ^{\prime }+\left( \overset{\cdot }{\xi }+\frac{1}{2}a\psi
\right) \xi ^{\prime }\right] \\ 
\text{ \ \ \ \ \ \ \ \ \ \ \ \ \ \ \ \ \ \ \ \ \ \ \ \ \ \ \ \ \ \ \ \ \ \ \
\ }+\frac{1}{2}r\left[ \left( \psi ^{\prime }\right) ^{2}+\left( \xi
^{\prime }\right) ^{2}\right] ,%
\end{array}
\tag{2.23}  \label{2.23}
\end{equation}%
\begin{equation}
\nu ^{\prime }-\lambda ^{\prime }+r^{-1}\left( 1-e^{2\lambda }\right)
+r\left( \frac{1}{2}e^{-2\nu }\left( a^{\prime }\right) ^{2}+e^{2\lambda
}V\right) =0,  \tag{2.24}  \label{2.24}
\end{equation}%
\begin{equation}
\lambda ^{\prime }+\nu ^{\prime }=r\left[ \left( \psi ^{\prime }\right)
^{2}+\left( \xi ^{\prime }\right) ^{2}\right] ,  \tag{2.25}  \label{2.25}
\end{equation}%
\begin{equation}
\begin{array}{l}
v^{\prime \prime }+\left( \nu ^{\prime }-\lambda ^{\prime }\right) \left(
\nu ^{\prime }+r^{-1}\right) -e^{\lambda -\nu }\left[ \left( \overset{\cdot }%
{\lambda }\right) ^{\prime }+\left( \overset{\cdot }{\nu }\right) ^{\prime
}+2\left( \overset{\cdot }{\psi }-\frac{1}{2}a\xi \right) \psi ^{\prime
}+2\left( \overset{\cdot }{\xi }+\frac{1}{2}a\psi \right) \xi ^{\prime }%
\right] \\ 
-\frac{1}{2}e^{-2\nu }\left( a^{\prime }\right) ^{2}+\left[ \left( \psi
^{\prime }\right) ^{2}+\left( \xi ^{\prime }\right) ^{2}\right] +e^{2\lambda
}V=0.%
\end{array}
\tag{2.26}  \label{2.26}
\end{equation}%
Using\ $\left( \ref{2.8}\right) $, $\left( \ref{2.14}\right) $ and $\left( %
\ref{2.15}\right) $ in\ $\left( \ref{2.5}\right) $, the relevant Yang-Mills
equations are the following%
\begin{equation}
a^{\prime \prime }+\left( 2r^{-1}-\lambda ^{\prime }-\nu ^{\prime }\right)
a^{\prime }-e^{\lambda -\nu }\left( \left( \overset{\cdot }{a}\right)
^{\prime }-\overset{\cdot }{\lambda }-\overset{\cdot }{\nu }\right)
=e^{2\lambda }\left[ \psi \overset{\cdot }{\xi }-\xi \overset{\cdot }{\psi }+%
\frac{1}{2}a\left( \psi ^{2}+\xi ^{2}\right) \right] ,  \tag{2.27}
\label{2.27}
\end{equation}%
\begin{equation}
a^{\prime \prime }+\left( 2r^{-1}-\lambda ^{\prime }-\nu ^{\prime }\right)
a^{\prime }=e^{\lambda +\nu }\left( \psi \xi ^{\prime }-\xi \psi ^{\prime
}\right) .  \tag{2.28}  \label{2.28}
\end{equation}%
Using\ $\left( \ref{2.4}\right) $, $\left( \ref{2.8}\right) $, $\left( \ref%
{2.9}\right) $ and $\left( \ref{2.10}\right) $ in\ $\left( \ref{2.5}\right) $%
, the Higgs equations are\ found\ to\ be\ equivalent\ to the following
system 
\begin{equation}
\begin{array}{c}
2\left( \overset{\cdot }{\psi }\right) ^{\prime }-a\xi ^{\prime }-\frac{1}{2}%
a^{\prime }\xi +2r^{-1}\left( \overset{\cdot }{\psi }-\frac{1}{2}a\xi
\right) -e^{\nu -\lambda }\left[ \psi ^{\prime \prime }+\psi ^{\prime
}\left( 2r^{-1}+\nu ^{\prime }-\lambda ^{\prime }\right) \right] =-\psi
V^{\prime }e^{\lambda +\nu }, \\ 
2\left( \overset{\cdot }{\xi }\right) ^{\prime }+a\psi ^{\prime }+\frac{1}{2}%
a^{\prime }\psi +2r^{-1}\left( \overset{\cdot }{\xi }+\frac{1}{2}a\psi
\right) -e^{\nu -\lambda }\left[ \xi ^{\prime \prime }+\xi ^{\prime }\left(
2r^{-1}+\nu ^{\prime }-\lambda ^{\prime }\right) \right] =-\xi V^{\prime
}e^{\lambda +\nu }.%
\end{array}
\tag{2.29}  \label{2.29}
\end{equation}

\begin{remark}
$\left( i\right) $ It\ is\ shown\ in \cite{25} that if the YMH equations are
satisfied, then the energy-momentum-stress tensor given by $\left( \ref{2.6}%
\right) $\ is divergence-free, i.e., $g^{\alpha \sigma }\nabla _{\sigma
}T_{\alpha \beta }=0$. Setting, as in \cite{6}, $E_{\alpha \beta }=R_{\alpha
\beta }-\frac{1}{2}Rg_{\alpha \beta }-T_{\alpha \beta }$, and using the
contracted Bianchi identity $g^{\alpha \sigma }\nabla _{\sigma }\left(
R_{\alpha \beta }-\frac{1}{2}Rg_{\alpha \beta }\right) =0$, it easily
follows that 
\begin{equation}
g^{\alpha \sigma }\nabla _{\sigma }E_{\alpha \beta }=0,  \tag{2.30}
\label{2.30}
\end{equation}%
provided that the YMH equations are satisfied.

$\left( ii\right) $ It is straightforward to check by using $\left( \ref%
{2.30}\right) $ that, if the Einstein equations $\left( \ref{2.24}\right) $
and $\left( \ref{2.25}\right) $, i.e., $E\left( e_{1},e_{1}\right) =0$ and $%
E\left( e_{0},e_{1}\right) =0$, are satisfied as well as the YMH equations $%
\left( \ref{2.27}\right) $, $\left( \ref{2.28}\right) $ and $\left( \ref%
{2.29}\right) $, then the Einstein equation $\left( \ref{2.26}\right) $ is
satisfied, i.e., $E\left( e_{2},e_{2}\right) =E\left( e_{3},e_{3}\right) =0$%
. Furthermore $E\left( e_{0},e_{0}\right) $ solves the following PDE 
\begin{equation*}
\left[ E\left( e_{0},e_{0}\right) \right] ^{\prime }+2\left( v^{\prime
}+r^{-1}\right) E\left( e_{0},e_{0}\right) =0,
\end{equation*}%
which yields 
\begin{equation*}
E\left( e_{0},e_{0}\right) \left( u,r\right) =\left( \frac{r_{0}}{r}\right)
^{2}E\left( e_{0},e_{0}\right) \left( u,r_{0}\right) \exp \left[ 2\left(
v\left( u,r_{0}\right) -v\left( u,r_{0}\right) \right) \right] .
\end{equation*}%
This implies $E\left( e_{0},e_{0}\right) =0$ if regularity is assumed at the
center.

$\left( iii\right) $ From the above observations we conclude that the
Einstein equations $\left( \ref{2.24}\right) $ and $\left( \ref{2.25}\right) 
$, together with the YMH equations $\left( \ref{2.27}\right) $, $\left( \ref%
{2.28}\right) $ and $\left( \ref{2.29}\right) $, are equivalent to the full
set of EYMH equations $\left( 2.23-2.29\right) $.
\end{remark}

\section{Reduction of the EYMH equations to a non linear evolution system}

We adapt the tools set up and implemented in \cite{6}. Two new functions are
introduced as follows:%
\begin{equation}
h=\left( r\psi \right) ^{\prime },\quad k=\left( r\xi \right) ^{\prime }. 
\tag{3.1}  \label{3.1}
\end{equation}%
Then%
\begin{equation}
\psi =\overline{h}=\frac{1}{r}\int_{0}^{r}h\left( s\right) ds,\quad \xi =%
\overline{k}=\frac{1}{r}\int_{0}^{r}k\left( s\right) ds,\quad \psi ^{\prime
}=\frac{h-\overline{h}}{r},\quad \xi ^{\prime }=\frac{k-\overline{k}}{r}. 
\tag{3.2}  \label{3.2}
\end{equation}%
The Einstein equation $\left( \ref{2.25}\right) $ then reads%
\begin{equation*}
\lambda ^{\prime }+\nu ^{\prime }=\frac{1}{r}\left[ \left( h-\overline{h}%
\right) ^{2}+\left( k-\overline{k}\right) ^{2}\right] ,
\end{equation*}%
and the solution, which satisfies the asymptotic condition $\lambda +\nu
\longrightarrow 0$ as $r\longrightarrow \infty $, is 
\begin{equation}
\lambda +\nu =-\int_{r}^{+\infty }\frac{1}{s}\left[ \left( h-\overline{h}%
\right) ^{2}+\left( k-\overline{k}\right) ^{2}\right] ds.  \tag{3.3}
\label{3.3}
\end{equation}%
Integrating the Yang-Mills equation $\left( \ref{2.28}\right) $ yields%
\begin{equation}
a\left( u,r\right) =\int_{0}^{r}e^{\lambda +\nu }\frac{Q\left( u,s\right) }{%
s^{2}}ds,  \tag{3.4}  \label{3.4}
\end{equation}%
where $Q$ is the local charge function defined by 
\begin{equation}
Q\left( u,r\right) =\int_{0}^{r}s\left( \overline{h}k-\overline{k}h\right)
\left( u,s\right) ds.  \tag{3.5}  \label{3.5}
\end{equation}%
Now we put the Einstein equation $\left( \ref{2.24}\right) $ under the form 
\begin{equation*}
\left( \nu ^{\prime }-\lambda ^{\prime }\right) e^{\nu -\lambda
}+r^{-1}e^{\nu -\lambda }=r^{-1}e^{\lambda +\nu }-r\left( \frac{1}{2}%
e^{-\lambda -\nu }\left( a^{\prime }\right) ^{2}+e^{\lambda +\nu }V\right) ,
\end{equation*}%
which is integrated to give%
\begin{equation}
e^{\nu -\lambda }=\frac{1}{r}\int_{0}^{r}e^{\lambda +\nu }\left[ 1-\frac{1}{2%
}\frac{Q^{2}}{s^{2}}-s^{2}V\right] ds.  \tag{3.6}  \label{3.6}
\end{equation}%
From $\left( \ref{2.29}\right) $ and $\left( \ref{3.1}\right) $\ we recast
the Higgs equations into the following non linear evolution system%
\begin{equation}
\begin{array}{c}
\overset{\cdot }{h}-\frac{1}{2}e^{\nu -\lambda }h^{\prime }=\frac{1}{2r}%
\left( e^{\lambda +\nu }-e^{\nu -\lambda }\right) \left( h-\overline{h}%
\right) -\frac{Q^{2}}{4r^{3}}e^{\lambda +\nu }\left( h-\overline{h}\right) +%
\frac{Q}{4r}\overline{k}e^{\lambda +\nu } \\ 
-\frac{1}{2}\left[ r\left( h-\overline{h}\right) Ve^{\lambda +\nu }+r%
\overline{h}V^{\prime }e^{\lambda +\nu }-ak\right] , \\ 
\overset{\cdot }{k}-\frac{1}{2}e^{\nu -\lambda }k^{\prime }=\frac{1}{2r}%
\left( e^{\lambda +\nu }-e^{\nu -\lambda }\right) \left( k-\overline{k}%
\right) -\frac{Q^{2}}{4r^{3}}e^{\lambda +\nu }\left( k-\overline{k}\right) +%
\frac{Q}{4r}\overline{h}e^{\lambda +\nu } \\ 
-\frac{1}{2}\left[ r\left( k-\overline{k}\right) Ve^{\lambda +\nu }+r%
\overline{k}V^{\prime }e^{\lambda +\nu }+ah\right] .%
\end{array}
\tag{3.7}  \label{3.7}
\end{equation}%
We write system $\left( \ref{3.7}\right) $ in matrix form with unknown
vector function $W=\left( 
\begin{array}{c}
h \\ 
k%
\end{array}%
\right) $\ as follows%
\begin{equation}
DW=\frac{1}{2r}\left( g-\widetilde{g}\right) \left( W-\overline{W}\right) -%
\frac{Q^{2}g}{4r^{3}}\left( W-\overline{W}\right) +\frac{Qg}{4r}\sigma _{1}%
\overline{W}+\frac{a}{2}i\sigma _{2}W-\frac{r}{2}\left[ Vg\left( W-\overline{%
W}\right) +V^{\prime }g\overline{W}\right] ,  \tag{3.8}  \label{3.8}
\end{equation}%
where%
\begin{equation}
D=\frac{\partial }{\partial u}-\frac{\widetilde{g}}{2}\frac{\partial }{%
\partial r},\quad g=e^{\lambda +\nu },\quad \widetilde{g}=e^{\nu -\lambda }.
\tag{3.9}  \label{3.9}
\end{equation}

\begin{remark}
$\left( i\right) $ It is worth emphasizing that the integration of the
Einstein equations is achieved under the asymptotic condition $\lambda +\nu
\rightarrow 0,$ $r\rightarrow \infty $. The Yang-Mills equation $\left( \ref%
{2.28}\right) $ is integrated easily by classical tools (change of unknown
function and variation of the constant) to yield the solution $a\left(
u,r\right) $ that exists for all $r\in \lbrack 0,+\infty )$.

$\left( ii\right) $ For the Yang-Mills field it is easy to see that it
vanishes at spatial infinity i.e., $F\left( u,r\right) \rightarrow 0,$ $%
r\rightarrow +\infty $. Actually, by simple calculation, all the components
of the Yang-Mills field $F$ vanish except $F_{01}$ which is given by $%
F_{01}=-a^{\prime }T_{3}$. Using the expression of $a\left( u,r\right) $ and
estimating the local charge $Q\left( u,r\right) $ one easily gains $%
F_{01}\left( u,r\right) \rightarrow 0,$ $r\rightarrow +\infty $.

$\left( iii\right) $ It could be of interest to investigate the case of an
asymptotic (anti) de Sitter metric.
\end{remark}

\section{Existence and uniqueness of global classical solutions}

This section aims at stating and proving a global existence and uniqueness
result for the initial value problem associated with the non linear
evolution system $\left( \ref{3.8}\right) $. In order to do this we begin by
introducing the spaces of functions used and some preliminary notations.

\subsection{Functional framework, notations, and statement of the main result%
}

For a vector function $W=\left( 
\begin{array}{c}
h \\ 
k%
\end{array}%
\right) $ defined on $[0,\infty )\times \lbrack 0,\infty )$ with the real
functions $h$ and $k$ belonging both to $C^{1}\left( [0,\infty )\times
\lbrack 0,\infty )\right) $, we will just write $W\in C^{1}\left( [0,\infty
)\times \lbrack 0,\infty )\right) $ instead of $W\in \left[ C^{1}\left(
[0,\infty )\times \lbrack 0,\infty )\right) \right] ^{2}$. We will also use
the notation $\left\vert W\right\vert :=\left\vert h\right\vert +\left\vert
k\right\vert $. Consider the initial value problem for system $\left( \ref%
{3.8}\right) $ with initial datum $W_{0}\left( r\right) =W\left( 0,r\right) $%
. Following the works in \cite{4} we define the Banach function spaces $%
\left( \mathcal{X},\left\Vert .\right\Vert _{\mathcal{X}}\right) ,$ $\left( 
\mathcal{X}_{0},\left\Vert .\right\Vert _{\mathcal{X}_{0}}\right) $, and $%
\left( \mathcal{Y},\left\Vert .\right\Vert _{\mathcal{Y}}\right) $\ by%
\begin{equation}
\begin{array}{l}
\mathcal{X}=\left\{ W=W\left( u,r\right) \in C^{1}\left( [0,\infty )\times
\lbrack 0,\infty )\right) :\left\Vert W\right\Vert _{\mathcal{X}}<\infty
\right\} , \\ 
\mathcal{X}_{0}=\left\{ v=v\left( r\right) \in C^{1}\left( [0,\infty
)\right) :\left\Vert v\right\Vert _{\mathcal{X}_{0}}<\infty \right\} , \\ 
\mathcal{Y}=\left\{ W=W\left( u,r\right) \in C^{1}\left( [0,\infty )\times
\lbrack 0,\infty )\right) :W\left( 0,r\right) =W_{0}\left( r\right) ,\quad
\left\Vert W\right\Vert _{\mathcal{Y}}<\infty \right\} ,%
\end{array}
\tag{4.1}  \label{4.1}
\end{equation}%
where%
\begin{equation*}
\begin{array}{l}
\left\Vert W\right\Vert _{\mathcal{X}}:=\underset{u,r\geq 0}{\sup }\left\{
\left( 1+r+u\right) ^{2}\left\vert W\left( u,r\right) \right\vert +\left(
1+r+u\right) ^{3}\left\vert W^{\prime }\left( u,r\right) \right\vert
\right\} , \\ 
\left\Vert v\right\Vert _{\mathcal{X}_{0}}:=\underset{r\geq 0}{\sup }\left\{
\left( 1+r\right) ^{2}\left\vert v\left( r\right) \right\vert +\left(
1+r\right) ^{3}\left\vert v^{\prime }\left( r\right) \right\vert \right\} ,
\\ 
\left\Vert W\right\Vert _{\mathcal{Y}}:=\underset{u,r\geq 0}{\sup }\left\{
\left( 1+r+u\right) ^{2}\left\vert W\left( u,r\right) \right\vert \right\} .%
\end{array}%
\end{equation*}%
We are now in the position to state the main result of this paper which is
the generalization, as we have mentioned from the outset, of the results
obtained in \cite{4, 6}.

\begin{theorem}
Assume for the self-interaction potential $V$ that $V\in C^{2}\left(
[0,\infty )\right) $, and there exists a constant $K_{0}\geq 0$ such that%
\begin{equation}
|V(t)|+t|V^{\prime }(t)|+t^{2}|V^{\prime \prime }(t)|\leq K_{0}t^{p+1}\quad
\forall t\geq 0\text{, for\ some\ }p\geq \frac{3}{2}.  \tag{4.2}  \label{4.2}
\end{equation}%
Suppose for the initial datum $W_{0}$ that 
\begin{equation}
W_{0}\in C^{1}([0,\infty )),\quad W_{0}\left( r\right) =O\left(
r^{-2}\right) \quad W_{0}^{\prime }\left( r\right) =O\left( r^{-3}\right) . 
\tag{4.3}  \label{4.3}
\end{equation}%
Then there exists $\epsilon >0$ such that if $\left\Vert W_{0}\right\Vert _{%
\mathcal{X}_{0}}<\epsilon $, then there exists a unique global classical
solution $W\in C^{1}\left( [0,\infty )\times \lbrack 0,\infty )\right) $ of $%
\left( \ref{3.8}\right) $ such that $W\left( 0,r\right) =W_{0}\left(
r\right) $. In addition this solution fulfills the decay property%
\begin{equation}
|W(u,r)|\leq C(1+u+r)^{-2},\quad |W^{\prime }(u,r)|\leq C(1+u+r)^{-3}, 
\tag{4.4}  \label{4.4}
\end{equation}%
where $C>0$\ is\ an\ increasing\ continuous\ function\ of $K_{0}$.\
Moreover, the corresponding space-time is time-like and null geodesically
complete toward the future.
\end{theorem}

\begin{remark}
$\left( i\right) $ Theorem 4.1 was stated and proved, under more restrictive
assumptions, by D.\ Chae \cite{4} for the Einstein-Maxwell-Higgs model.
Notably, in \cite{4}, the assumption on $p$ was $p\geq 3$. It is our aim, in
this paper, to provide a generalization and improvement of the results
obtained in \cite{4, 6} by extending them, under assumption $\left( \ref{4.2}%
\right) $,\ to the EYMH model. Note also that, unlike the solution obtained
by D. Christodoulou \cite{6} for the spherically symmetric Einstein-Scalar
field system, the solution obtained in the present work decays more\ slowly
than that of \cite{6}. This latter fact is attributed, as\ it will\ be\
shown thereafter, to the non-vanishing of the local charge $Q$.

$\left( ii\right) $ Note that assumption $\left( \ref{4.2}\right) $ is\
satisfied for $V\equiv 0$. Theorem 4.1 then applies to provide global
existence and uniqueness of classical solutions of\ the\ spherically
symmetric EYMH system with\ vanishing\ self-interaction\ potential.

$\left( ii\right) $ \textit{If }$\Phi $ is a real doublet of the form $\Phi
=\left( 
\begin{array}{c}
0 \\ 
\psi%
\end{array}%
\right) $, i.e., $\xi \equiv 0$, then the local charge $Q$ as well as the
Yang-Mills potential $A$ and the Yang-Mills strength field $F$ vanish (see $%
\left( \ref{3.1}\right) ,$ $\left( \ref{3.5}\right) ,$ $\left( \ref{3.4}%
\right) $\ and $\left( \ref{2.14}\right) $). In this case we can choose $%
V\left( t\right) =\frac{t^{p+1}}{p+1}$, $t\geq 0$, so that the Higgs
equations reads%
\begin{equation}
g^{\lambda \mu }\nabla _{\lambda }\nabla _{\mu }\psi =\psi ^{2p+1}. 
\tag{4.5}  \label{4.5}
\end{equation}%
$\left( \ref{4.5}\right) $ is the non linear Klein-Gordon equation that can
be reduced to the form%
\begin{equation*}
Dh=\frac{1}{2r}\left( g-\widetilde{g}\right) \left( h-\overline{h}\right) -%
\frac{rg}{2}\left[ \left( h-\overline{h}\right) \frac{\left( \overline{h}%
\right) ^{2p+2}}{p+1}+\left( \overline{h}\right) ^{2p+1}\right] ,
\end{equation*}%
where 
\begin{equation*}
\widetilde{g}=\frac{1}{r}\int_{0}^{r}g\left[ 1-s^{2}\frac{\left( \overline{h}%
\right) ^{2p+2}}{p+1}\right] ds.
\end{equation*}%
Theorem 4.1 then applies in this situation to encompass global existence and
uniqueness of classical spherically symmetric solutions of the non linear
Einstein-Klein-Gordon system. Moreover, in this case, it turns out that the
solutions obtained possess the same order of decay estimates as those
obtained by Christodoulou \cite{6}. The same statement holds true if $\Phi $
is of the form $\Phi =\left( 
\begin{array}{c}
0 \\ 
i\xi%
\end{array}%
\right) $, i.e., $\psi \equiv 0$.
\end{remark}

\subsection{Proof of Theorem 4.1}

Throughout this paragraph, given that $\left\{ I_{2},\sigma _{1},i\sigma
_{2},\sigma _{3}\right\} $ is a basis for the real vector space of $2\times
2 $ real matrices, $I_{2}$ being the $2\times 2$ identity matrix, we will
use the notation 
\begin{equation*}
\left\vert N\right\vert =\left\vert n_{0}\right\vert +\left\vert
n_{1}\right\vert +\left\vert n_{2}\right\vert +\left\vert n_{3}\right\vert ,
\end{equation*}%
if $N$ is a $2\times 2$ real matrix with 
\begin{equation*}
N=n_{0}I_{2}+n_{1}\sigma _{1}+n_{2}\left( i\sigma _{2}\right) +n_{3}\sigma
_{3}.
\end{equation*}%
We prove Theorem 4.1 by a contraction mapping argument as in \cite{4, 6, 24}
whilst improving and correcting at the same time some key estimates therein.
We define the mapping $\mathcal{K}:W\rightarrow w=\mathcal{K}(W)$, where $w$
is the solution of the first order linear initial value problem 
\begin{equation}
\begin{array}{l}
Dw=\frac{1}{2r}\left( g-\widetilde{g}\right) \left( w-\overline{W}\right) -%
\frac{Q^{2}g}{4r^{3}}\left( w-\overline{W}\right) +\frac{Qg}{4r}\sigma _{1}%
\overline{W}+\frac{a}{2}i\sigma _{2}w-\frac{r}{2}\left[ Vg\left( w-\overline{%
W}\right) +V^{\prime }g\overline{W}\right] , \\ 
w\left( 0,r\right) =W_{0}\left( r\right) .%
\end{array}
\tag{4.6}  \label{4.6}
\end{equation}%
Our purpose is to show that the mapping $\mathcal{K}$ defined above is a
contraction from a non-empty closed ball of $\mathcal{X}$ into itself. If
this is done, then the standard fixed point theorem applies to yield the
unique fixed point $W\in \mathcal{X}$ such that $\mathcal{K}(W)=W$, which is
the solution of the nonlinear system $\left( \ref{3.8}\right) $ satisfying $%
W\left( 0,r\right) =W_{0}\left( r\right) $.

\subsubsection{$\mathcal{K}$ is a mapping from a ball of $\mathcal{X}$ into
itself}

Let $\rho >0$ be a real number. $B_{\rho }$ denotes the closed ball, in $%
\mathcal{X}$, of radius $\rho $ centered at $0$ i.e., 
\begin{equation*}
B_{\rho }=\left\{ W\in \mathcal{X}:\left\Vert W\right\Vert _{\mathcal{X}%
}\leq \rho \right\} .
\end{equation*}%
We will prove that $\rho $ can be chosen small enough such that $\mathcal{K}%
:B_{\rho }\longrightarrow B_{\rho }.$ Let $W\in B_{\rho }$. We have to
estimate $\left\Vert \mathcal{K}\left( W\right) \right\Vert _{\mathcal{X}}$
in terms of $\left\Vert W\right\Vert _{\mathcal{X}}$. The characteristic
system of ODE associated to the initial value problem $\left( \ref{4.6}%
\right) $ is 
\begin{equation}
\begin{array}{l}
\frac{dr}{du}=-\frac{1}{2}\widetilde{g}, \\ 
\frac{dw}{du}=\frac{1}{2r}\left( g-\widetilde{g}\right) \left( w-\overline{W}%
\right) -\frac{Q^{2}g}{4r^{3}}\left( w-\overline{W}\right) +\frac{Qg}{4r}%
\sigma _{1}\overline{W}+\frac{a}{2}i\sigma _{2}w-\frac{r}{2}\left[ Vg\left(
w-\overline{W}\right) +V^{\prime }g\overline{W}\right] ,%
\end{array}
\tag{4.7}  \label{4.7}
\end{equation}%
with initial data 
\begin{equation*}
r\left( 0\right) =r_{0},\quad w\left( 0\right) =W_{0}.
\end{equation*}%
Let $r\left( u\right) =\gamma \left( u,r_{0}\right) $ be the solution of the
initial value problem 
\begin{equation}
\frac{dr}{du}=-\frac{1}{2}\widetilde{g}\left( u,r\right) ,\quad r\left(
0\right) =r_{0}.  \tag{4.8}  \label{4.8}
\end{equation}%
Then 
\begin{equation}
r_{1}=r_{0}-\frac{1}{2}\int_{0}^{u_{1}}\widetilde{g}\left( u,\gamma \left(
u,r_{0}\right) \right) du,  \tag{4.9}  \label{4.9}
\end{equation}%
where $r_{1}=\gamma \left( u_{1},r_{0}\right) $. Integrating the second ODE
of $\left( \ref{4.7}\right) $\ along $\gamma $ we obtain 
\begin{equation}
w\left( u_{1},r_{1}\right) =\exp \left( \int_{0}^{u_{1}}\left[ N\left(
u,r\right) \right] _{\gamma }dv\right) W\left( 0,r_{0}\right)
+\int_{0}^{u_{1}}\left\{ \exp \left( \int_{u}^{u_{1}}\left[ N\left(
u,r\right) \right] _{\gamma }dv\right) \right\} \left[ f\right] _{\gamma }du,
\tag{4.10}  \label{4.10}
\end{equation}%
where the matrix function $N$ and the vector function $f$ are given by%
\begin{equation}
\begin{array}{l}
N\left( u,r\right) =\left( \frac{1}{2r}\left( g-\widetilde{g}\right) -\frac{%
Q^{2}g}{4r^{3}}-\frac{r}{2}Vg\right) I_{2}+\frac{a}{2}i\sigma _{2}, \\ 
f\left( u,r\right) =\left( \frac{Q^{2}g}{4r^{3}}-\frac{1}{2r}\left( g-%
\widetilde{g}\right) +\frac{rg}{2}\left[ V-V^{\prime }\right] \right) 
\overline{W}+\frac{Qg}{4r}\sigma _{1}\overline{W}.%
\end{array}
\tag{4.11}  \label{4.11}
\end{equation}%
We have to estimate $N$\ and $f$. Setting $\left\Vert W\right\Vert _{%
\mathcal{X}}=x$, we derive the following estimates as in \cite{4}%
\begin{equation}
\left\vert \sigma _{1}\overline{W}\left( u,r\right) \right\vert =\left\vert 
\overline{W}\left( u,r\right) \right\vert \leq \frac{x}{\left( 1+u\right)
\left( 1+u+r\right) },  \tag{4.12}  \label{4.12}
\end{equation}%
\begin{equation}
\left\vert W\left( u,r\right) -\overline{W}\left( u,r\right) \right\vert
\leq \frac{xr}{2\left( 1+u\right) \left( 1+u+r\right) ^{2}}.  \tag{4.13}
\label{4.13}
\end{equation}%
$\left( \ref{4.13}\right) $\ implies%
\begin{equation}
\int_{0}^{\infty }\frac{\left\vert W\left( u,r\right) -\overline{W}\left(
u,r\right) \right\vert ^{2}}{r}dr\leq \frac{x^{2}}{24\left( 1+u\right) ^{4}}.
\tag{4.14}  \label{4.14}
\end{equation}%
In\ view of $\left( \ref{3.3}\right) $ and the\ definition\ of $g$\ in $%
\left( \ref{3.9}\right) $, $\left( \ref{4.14}\right) $ yields%
\begin{equation}
g\left( u,0\right) \geq \exp \left( -\frac{x^{2}}{24\left( 1+u\right) ^{4}}%
\right) .  \tag{4.15}  \label{4.15}
\end{equation}%
From $\left( \ref{4.13}\right) $ we gain 
\begin{equation}
\begin{array}{l}
\left\vert g\left( u,r\right) -\overline{g}\left( u,r\right) \right\vert
\leq \frac{1}{r}\int_{0}^{r}\left\vert g\left( u,r\right) -g\left(
u,r^{\prime }\right) \right\vert dr^{\prime } \\ 
\text{ \ \ \ \ \ \ \ \ \ \ \ \ \ \ \ \ \ \ \ \ \ \ \ \ \ \ }\leq \frac{1}{r}%
\int_{0}^{r}\left[ \int_{r^{\prime }}^{r}\left\vert \frac{\partial g\left(
u,s\right) }{\partial s}\right\vert ds\right] dr^{\prime } \\ 
\text{ \ \ \ \ \ \ \ \ \ \ \ \ \ \ \ \ \ \ \ \ \ \ \ \ \ \ \ }\leq \frac{1}{r%
}\int_{0}^{r}\left[ \int_{r^{\prime }}^{r}\frac{\left\vert W\left(
u,s\right) -\overline{W}\left( u,s\right) \right\vert ^{2}}{s}ds\right]
dr^{\prime } \\ 
\text{ \ \ \ \ \ \ \ \ \ \ \ \ \ \ \ \ \ \ \ \ \ \ \ \ \ \ \ }\leq \frac{%
x^{2}}{4r\left( 1+u\right) ^{2}}\int_{0}^{r}\left[ \int_{r^{\prime }}^{r}%
\frac{s}{\left( 1+u+s\right) ^{4}}ds\right] dr^{\prime } \\ 
\text{ \ \ \ \ \ \ \ \ \ \ \ \ \ \ \ \ \ \ \ \ \ \ \ \ \ \ \ }=\frac{%
x^{2}r^{2}}{12\left( 1+u\right) ^{3}\left( 1+u+r\right) ^{3}}.%
\end{array}
\tag{4.16}  \label{4.16}
\end{equation}%
In view of $\left( \ref{3.5}\right) $\ and\ $\left( \ref{4.12}\right) $\ we
estimate the local charge to get 
\begin{equation}
\begin{array}{l}
\left\vert Q\left( u,r\right) \right\vert \leq \int_{0}^{r}\left\vert
W\left( u,s\right) \right\vert \left\vert \overline{W}\left( u,s\right)
\right\vert sds \\ 
\text{ \ \ \ \ \ \ \ \ \ \ \ \ }\leq \frac{x^{2}}{1+u}\int_{0}^{r}\frac{s}{%
\left( 1+u+s\right) ^{3}}ds \\ 
\text{ \ \ \ \ \ \ \ \ \ \ \ \ }=\frac{x^{2}r^{2}}{2\left( 1+u\right)
^{2}\left( 1+u+r\right) ^{2}}.%
\end{array}
\tag{4.17}  \label{4.17}
\end{equation}%
Hence%
\begin{equation}
\frac{1}{r}\int_{0}^{r}\frac{\left\vert Q\left( u,s\right) \right\vert ^{2}}{%
s^{2}}ds\leq \frac{x^{4}}{4r\left( 1+u\right) ^{4}}\int_{0}^{r}\frac{s^{2}}{%
\left( 1+u+s\right) ^{4}}ds=\frac{x^{4}r^{2}}{12\left( 1+u\right) ^{5}\left(
1+u+r\right) ^{3}}.  \tag{4.18}  \label{4.18}
\end{equation}%
The term containing\ the\ self-interaction\ potential\ is estimated using $%
\Phi ^{\dag }\Phi =\left( \overline{h}\right) ^{2}+\left( \overline{k}%
\right) ^{2}$ and assumption $\left( \ref{4.2}\right) $\ as follows%
\begin{equation}
\begin{array}{l}
\left\vert \frac{1}{r}\int_{0}^{r}s^{2}V\left( \Phi ^{\dag }\Phi \right)
ds\right\vert \leq \frac{K_{0}}{r}\int_{0}^{r}s^{2}\left\vert \overline{W}%
\left( u,s\right) \right\vert ^{2p+2}ds \\ 
\text{ \ \ \ \ \ \ \ \ \ \ \ \ \ \ \ \ \ \ \ \ \ \ \ \ \ \ \ \ \ \ \ }\leq 
\frac{K_{0}}{r}\frac{x^{2p+2}}{\left( 1+u\right) ^{2p+2}}\int_{0}^{r}\frac{%
s^{2}}{\left( 1+u+s\right) ^{2p+2}}ds \\ 
\text{ \ \ \ \ \ \ \ \ \ \ \ \ \ \ \ \ \ \ \ \ \ \ \ \ \ \ \ \ \ \ \ }\leq 
\frac{K_{0}x^{2p+2}r^{2}}{3\left( 1+u\right) ^{2p+3}\left( 1+u+r\right)
^{2p+1}} \\ 
\text{ \ \ \ \ \ \ \ \ \ \ \ \ \ \ \ \ \ \ \ \ \ \ \ \ \ \ \ \ \ \ \ }\leq 
\frac{K_{0}x^{2p+2}r^{2}}{3\left( 1+u\right) ^{6}\left( 1+u+r\right) ^{4}}.%
\end{array}
\tag{4.19}  \label{4.19}
\end{equation}

From the\ definition\ of $\widetilde{g}$\ in $\left( \ref{3.9}\right) $ and
the\ estimates $\left( \ref{4.16}\right) $, $\left( \ref{4.18}\right) $, and 
$\left( \ref{4.19}\right) $\ we get 
\begin{equation}
\begin{array}{l}
\left\vert g\left( u,r\right) -\widetilde{g}\left( u,r\right) \right\vert
\leq \left\vert g\left( u,r\right) -\overline{g}\left( u,r\right)
\right\vert +\frac{1}{2r}\int_{0}^{r}\frac{\left\vert Q\left( u,s\right)
\right\vert ^{2}}{s^{2}}ds+\frac{1}{r}\int_{0}^{r}s^{2}\left\vert V\left(
\Phi ^{\dag }\Phi \right) \right\vert ds \\ 
\text{ \ \ \ \ \ \ \ \ \ \ \ \ \ \ \ \ \ \ \ \ \ \ \ \ \ \ }\leq \frac{%
x^{2}r^{2}}{12\left( 1+u\right) ^{5}\left( 1+u+r\right) ^{3}}+\frac{%
x^{4}r^{2}}{24\left( 1+u\right) ^{5}\left( 1+u+r\right) ^{3}}+\frac{%
K_{0}x^{2p+2}r^{2}}{3\left( 1+u\right) ^{11}\left( 1+u+r\right) ^{4}} \\ 
\text{ \ \ \ \ \ \ \ \ \ \ \ \ \ \ \ \ \ \ \ \ \ \ \ \ \ \ }\leq \frac{%
\left( 3+8K_{0}\right) \left( x^{2}+x^{4}+x^{2p+2}\right) r^{2}}{24\left(
1+u\right) ^{5}\left( 1+u+r\right) ^{3}}.%
\end{array}
\tag{4.20}  \label{4.20}
\end{equation}%
In view of $\left( \ref{4.15}\right) $, $\left( \ref{4.16}\right) $, and $%
\left( \ref{4.20}\right) $ we find out that 
\begin{equation}
\begin{array}{l}
\widetilde{g}\left( u,r\right) \geq \overline{g}\left( u,0\right) -\frac{1}{%
2r}\int_{0}^{r}\frac{\left\vert Q\left( u,s\right) \right\vert ^{2}}{s^{2}}%
ds-\frac{1}{r}\int_{0}^{r}s^{2}V\left( \Phi ^{\dag }\Phi \right) ds \\ 
\text{ \ \ \ \ \ \ \ \ \ }\geq \exp \left( -\frac{x^{2}}{24}\right) -\frac{%
x^{4}}{24}-\frac{K_{0}x^{2p+2}}{3}.%
\end{array}
\tag{4.21}  \label{4.21}
\end{equation}%
It is easy to see that the function $l=l\left( x\right) =\exp \left( -\frac{%
x^{2}}{24}\right) -\frac{x^{4}}{24}-\frac{K_{0}x^{2p+2}}{3}$ has a unique
positive root $x_{0}$ and $l\left( x\right) \in (0,1]$ for all $x\in \lbrack
0,x_{0})$. We now estimate the function $f$ defined in $\left( \ref{4.11}%
\right) $ as\ follows.%
\begin{equation}
\left\vert f\left( u,r\right) \right\vert \leq \left\vert \frac{Q^{2}g%
\overline{W}}{4r^{3}}\right\vert +\left\vert \frac{1}{2r}\left( g-\widetilde{%
g}\right) \overline{W}\right\vert +\left\vert \frac{rg}{2}\left( V-V^{\prime
}\right) \overline{W}\right\vert +\left\vert \frac{Qg}{4r}\sigma _{1}%
\overline{W}\right\vert .  \tag{4.22}  \label{4.22}
\end{equation}%
Since $0<g\leq 1$,$\ \left( \ref{4.12}\right) $ and $\left( \ref{4.17}%
\right) $\ give%
\begin{equation}
\left\vert \frac{Q^{2}g\overline{W}}{4r^{3}}\right\vert \leq \frac{x^{5}r}{%
16\left( 1+u\right) ^{5}\left( 1+u+r\right) ^{5}},  \tag{4.23}  \label{4.23}
\end{equation}%
and%
\begin{equation}
\left\vert \frac{Qg}{4r}\sigma _{1}\overline{W}\right\vert \leq \frac{x^{3}r%
}{8\left( 1+u\right) ^{3}\left( 1+u+r\right) ^{3}}.  \tag{4.24}  \label{4.24}
\end{equation}%
Using $\left( \ref{4.12}\right) $ and $\left( \ref{4.20}\right) $ we gain%
\begin{equation}
\left\vert \frac{1}{2r}\left( g-\widetilde{g}\right) \overline{W}\right\vert
\leq \frac{\left( 3+8K_{0}\right) \left( x^{3}+x^{5}+x^{2p+3}\right) r}{%
48\left( 1+u\right) ^{6}\left( 1+u+r\right) ^{4}}.  \tag{4.25}  \label{4.25}
\end{equation}%
Assumption $\left( \ref{4.2}\right) $\ and\ the\ estimate\ $\left( \ref{4.12}%
\right) $\ imply 
\begin{equation}
\begin{array}{l}
\left\vert \frac{rg}{2}\left( V\left( \Phi ^{\dag }\Phi \right) -V^{\prime
}\left( \Phi ^{\dag }\Phi \right) \right) \overline{W}\right\vert \leq \frac{%
K_{0}rg}{2}\left( \left\vert \overline{W}\right\vert ^{2p+3}+\left\vert 
\overline{W}\right\vert ^{2p+1}\right) \\ 
\text{ \ \ \ \ \ \ \ \ \ \ \ \ \ \ \ \ \ \ \ \ \ \ \ \ \ \ \ \ \ \ \ \ \ \ \
\ \ \ \ \ \ \ \ \ \ \ \ \ \ \ \ }\leq \frac{K_{0}r}{2}\left[ \frac{x^{2p+3}}{%
\left( 1+u\right) ^{2p+3}\left( 1+u+r\right) ^{2p+3}}+\frac{x^{2p+1}}{\left(
1+u\right) ^{2p+1}\left( 1+u+r\right) ^{2p+1}}\right] \\ 
\text{ \ \ \ \ \ \ \ \ \ \ \ \ \ \ \ \ \ \ \ \ \ \ \ \ \ \ \ \ \ \ \ \ \ \ \
\ \ \ \ \ \ \ \ \ \ \ \ \ \ \ \ }\leq \frac{K_{0}}{2}\left[ \frac{x^{2p+3}}{%
\left( 1+u\right) ^{2p+3}\left( 1+u+r\right) ^{2p+2}}+\frac{x^{2p+1}}{\left(
1+u\right) ^{2p+1}\left( 1+u+r\right) ^{2p}}\right] \\ 
\text{ \ \ \ \ \ \ \ \ \ \ \ \ \ \ \ \ \ \ \ \ \ \ \ \ \ \ \ \ \ \ \ \ \ \ \
\ \ \ \ \ \ \ \ \ \ \ \ \ \ \ \ }\leq \frac{K_{0}}{2}\left[ \frac{x^{2p+3}}{%
\left( 1+u\right) ^{6}\left( 1+u+r\right) ^{5}}+\frac{x^{2p+1}}{\left(
1+u\right) ^{4}\left( 1+u+r\right) ^{3}}\right]%
\end{array}
\tag{4.26}  \label{4.26}
\end{equation}%
Inserting $\left( \ref{4.23}\right) $, $\left( \ref{4.24}\right) $, $\left( %
\ref{4.25}\right) $, and\ $\left( \ref{4.26}\right) $\ into\ $\left( \ref%
{4.22}\right) $\ yields%
\begin{equation}
\left\vert f\left( u,r\right) \right\vert \leq \frac{\left(
12+56K_{0}\right) \left( x^{3}+x^{5}+x^{2p+1}+x^{2p+3}\right) }{48\left(
1+u\right) ^{3}\left( 1+u+r\right) ^{2}}.  \tag{4.27}  \label{4.27}
\end{equation}%
For $r\left( u\right) =\gamma \left( u,r_{0}\right) $ it holds in view of $%
\left( \ref{4.8}\right) $\ and $\left( \ref{4.21}\right) $\ that 
\begin{equation}
r=r_{1}+\frac{1}{2}\int_{u}^{u_{1}}\widetilde{g}\left( s,r\left( s\right)
\right) ds\geq r_{1}+\frac{l\left( x\right) }{2}\left( u_{1}-u\right) . 
\tag{4.28}  \label{4.28}
\end{equation}%
Since $l\left( x\right) \in (0,1]$\ for all $x\in \lbrack 0,x_{0})$, $\left( %
\ref{4.28}\right) $ implies\ 
\begin{equation}
1+u+r\geq 1+u+r_{1}+\frac{l\left( x\right) }{2}\left( u_{1}-u\right) \geq 
\frac{l\left( x\right) }{2}\left( 1+u_{1}+r_{1}\right) .  \tag{4.29}
\label{4.29}
\end{equation}%
From $\left( \ref{4.27}\right) $ and $\left( \ref{4.29}\right) $ we gain the
estimate%
\begin{equation}
\begin{array}{l}
\int_{0}^{u_{1}}\left\vert \left[ f\right] _{\gamma }\right\vert
du=\int_{0}^{u_{1}}\left\vert f\left( u_{1},r_{1}\right) \right\vert du \\ 
\text{ \ \ \ \ \ \ \ \ \ \ \ \ \ \ \ \ \ }\leq \int_{0}^{u_{1}}\frac{\left(
12+56K_{0}\right) \left( x^{3}+x^{5}+x^{2p+1}+x^{2p+3}\right) }{48\left(
1+u_{1}\right) ^{3}\left( 1+u_{1}+r_{1}\right) ^{2}}du \\ 
\text{ \ \ \ \ \ \ \ \ \ \ \ \ \ \ \ \ \ }\leq \frac{\left(
12+56K_{0}\right) \left( x^{3}+x^{5}+x^{2p+1}+x^{2p+3}\right) }{48}%
\int_{0}^{u_{1}}\frac{1}{\left( 1+u\right) ^{3}\left( 1+u+r\right) ^{2}}du
\\ 
\text{ \ \ \ \ \ \ \ \ \ \ \ \ \ \ \ \ \ }\leq \frac{\left(
12+56K_{0}\right) \left( x^{3}+x^{5}+x^{2p+1}+x^{2p+3}\right) }{24\left(
1+u_{1}+r_{1}\right) ^{2}l^{2}\left( x\right) }\int_{0}^{\infty }\frac{1}{%
\left( 1+u\right) ^{3}}du \\ 
\text{ \ \ \ \ \ \ \ \ \ \ \ \ \ \ \ \ \ }=\frac{\left( 12+56K_{0}\right)
\left( x^{3}+x^{5}+x^{2p+1}+x^{2p+3}\right) }{48\left( 1+u_{1}+r_{1}\right)
^{2}l^{2}\left( x\right) }.%
\end{array}
\tag{4.30}  \label{4.30}
\end{equation}%
Let us now estimate the matrix function $N$ defined in $\left( \ref{4.11}%
\right) $ as\ follows%
\begin{equation}
\left\vert N\left( u,r\right) \right\vert \leq \left\vert \frac{1}{2r}\left(
g-\widetilde{g}\right) \right\vert +\left\vert \frac{Q^{2}g}{4r^{3}}%
\right\vert +\left\vert \frac{r}{2}Vg\right\vert +\left\vert \frac{a}{2}%
\right\vert .  \tag{4.31}  \label{4.31}
\end{equation}%
In view of $\left( \ref{4.20}\right) $ it holds that%
\begin{equation}
\left\vert \frac{1}{2r}\left( g-\widetilde{g}\right) \left( u,r\right)
\right\vert \leq \frac{\left( 3+8K_{0}\right) \left(
x^{2}+x^{4}+x^{2p+2}\right) }{48\left( 1+u\right) ^{5}}.  \tag{4.32}
\label{4.32}
\end{equation}%
Since $0<g\leq 1$, $\left( \ref{4.17}\right) $ yields%
\begin{equation}
\left\vert \frac{Q^{2}g}{4r^{3}}\left( u,r\right) \right\vert \leq \frac{%
x^{4}}{16\left( 1+u\right) ^{4}}.  \tag{4.33}  \label{4.33}
\end{equation}%
From assumption $\left( \ref{4.2}\right) $, since $p\geq \frac{3}{2}$ and\ $%
0<g\leq 1$, we have as in\ $\left( \ref{4.26}\right) $%
\begin{equation}
\begin{array}{l}
\left\vert \frac{rg\left( u,r\right) }{2}V\left( \Phi ^{\dag }\Phi \right)
\right\vert \leq \frac{K_{0}r\left\vert \overline{W}\left( u,r\right)
\right\vert ^{2p+2}}{2} \\ 
\text{ \ \ \ \ \ \ \ \ \ \ \ \ \ \ \ \ \ \ \ \ \ \ \ \ \ \thinspace
\thinspace\ \ \ }\leq \frac{K_{0}rx^{2p+2}}{2\left( 1+u\right) ^{2p+2}\left(
1+u+r\right) ^{2p+2}}\leq \frac{K_{0}x^{2p+2}}{2\left( 1+u\right) ^{5}}.%
\end{array}
\tag{4.34}  \label{4.34}
\end{equation}%
Using $\left( \ref{3.4}\right) $\ and $\left( \ref{4.17}\right) $\ we
estimate $a$ as follows%
\begin{equation}
\begin{array}{l}
\left\vert \frac{a\left( u,r\right) }{2}\right\vert =\left\vert \frac{1}{2}%
\int_{0}^{r}g\frac{Q\left( u,s\right) }{s^{2}}ds\right\vert \\ 
\text{ \ \ \ \ \ \ \ \ \ \ \ }\leq \frac{x^{2}}{2\left( 1+u\right) ^{2}}%
\int_{0}^{r}\frac{1}{\left( 1+u+s\right) ^{2}}ds \\ 
\text{ \ \ \ \ \ \ \ \ \ \ \ }\leq \frac{x^{2}}{2\left( 1+u\right) ^{2}}%
\int_{0}^{r}\frac{1}{\left( 1+u+s\right) ^{2}}ds \\ 
\text{ \ \ \ \ \ \ \ \ \ \ \ }=\frac{x^{2}r}{2\left( 1+u\right) ^{3}\left(
1+u+r\right) }\leq \frac{x^{2}}{2\left( 1+u\right) ^{3}}.%
\end{array}
\tag{4.35}  \label{4.35}
\end{equation}%
Inserting $\left( \ref{4.32}\right) $, $\left( \ref{4.33}\right) $, $\left( %
\ref{4.34}\right) $, and\ $\left( \ref{4.35}\right) $\ into\ $\left( \ref%
{4.31}\right) $\ yields 
\begin{equation}
\left\vert N\left( u,r\right) \right\vert \leq \frac{\left(
30+32K_{0}\right) \left( x^{2}+x^{4}+x^{2p+2}\right) }{48}.  \tag{4.36}
\label{4.36}
\end{equation}%
From $\left( \ref{4.32}\right) $ we get

\begin{equation}
\begin{array}{l}
\int_{0}^{u_{1}}\left\vert \left[ \frac{1}{2r}\left( g-\widetilde{g}\right) %
\right] _{\gamma }\right\vert du\leq \int_{0}^{u_{1}}\frac{\left(
3+8K_{0}\right) \left( x^{2}+x^{4}+x^{2p+2}\right) }{48\left( 1+u_{1}\right)
^{5}}du \\ 
\text{ \ \ \ \ \ \ \ \ \ \ \ \ \ \ \ \ \ \ \ \ \ \ \ \ \ \ \ \ \ \ \ \ \ \ }%
\leq \frac{\left( 3+8K_{0}\right) \left( x^{2}+x^{4}+x^{2p+2}\right) }{48}%
\int_{0}^{\infty }\frac{1}{\left( 1+u\right) ^{5}}du \\ 
\text{ \ \ \ \ \ \ \ \ \ \ \ \ \ \ \ \ \ \ \ \ \ \ \ \ \ \ \ \ \ \ \ \ \ \ }=%
\frac{\left( 3+8K_{0}\right) \left( x^{2}+x^{4}+x^{2p+2}\right) }{192}.%
\end{array}
\tag{4.37}  \label{4.37}
\end{equation}%
$\left( \ref{4.33}\right) $ yields%
\begin{equation}
\int_{0}^{u_{1}}\left\vert \left[ \frac{Q^{2}g}{4r^{3}}\right] _{\gamma
}\right\vert du\leq \int_{0}^{u_{1}}\frac{x^{4}}{16\left( 1+u_{1}\right) ^{4}%
}du\leq \frac{x^{4}}{16}\int_{0}^{\infty }\frac{1}{\left( 1+u\right) ^{4}}du=%
\frac{x^{4}}{48}.  \tag{4.38}  \label{4.38}
\end{equation}%
From $\left( \ref{4.34}\right) $\ we have 
\begin{equation}
\int_{0}^{u_{1}}\left\vert \left[ \frac{rg}{2}V\right] _{\gamma }\right\vert
du\leq \int_{0}^{u_{1}}\frac{K_{0}x^{2p+2}}{2\left( 1+u_{1}\right) ^{5}}%
du\leq \frac{K_{0}x^{2p+2}}{2}\int_{0}^{\infty }\frac{1}{\left( 1+u\right)
^{5}}du=\frac{K_{0}x^{2p+2}}{8}.  \tag{4.39}  \label{4.39}
\end{equation}%
Using $\left( \ref{4.35}\right) $ we gain 
\begin{equation}
\int_{0}^{u_{1}}\left\vert \left[ \frac{a}{2}\right] _{\gamma }\right\vert
du\leq x^{2}\int_{0}^{u_{1}}\frac{1}{2\left( 1+u_{1}\right) ^{3}}du\leq 
\frac{x^{2}}{2}\int_{0}^{\infty }\frac{1}{\left( 1+u\right) ^{3}}du=\frac{%
x^{2}}{4}.  \tag{4.40}  \label{4.40}
\end{equation}%
It follows from $\left( \ref{4.31}\right) $, $\left( \ref{4.37}\right) $, $%
\left( \ref{4.38}\right) $, $\left( \ref{4.39}\right) $, and\ $\left( \ref%
{4.40}\right) $\ that 
\begin{equation}
\left\vert \int_{0}^{u_{1}}\left[ N\right] _{\gamma }dv\right\vert \leq
K_{1}\left( x^{2}+x^{4}+x^{2p+2}\right) ,  \tag{4.41}  \label{4.41}
\end{equation}%
where 
\begin{equation*}
K_{1}=\frac{55+32K_{0}}{192}.
\end{equation*}%
Using $\left( \ref{4.9}\right) $\ it holds that, for $x\in \lbrack 0,x_{0})$,%
\begin{equation}
1+r_{0}=1+r_{1}+\frac{1}{2}\int_{0}^{u_{1}}\widetilde{g}\left( s,r\left(
s\right) \right) ds\geq 1+r_{1}+\frac{l\left( x\right) }{2}u_{1}\geq \frac{%
l\left( x\right) }{2}\left( 1+r_{1}+u_{1}\right) .  \tag{4.42}  \label{4.42}
\end{equation}%
Thus, by the definition of the Banach space $\left( \mathcal{X}%
_{0},\left\Vert .\right\Vert _{\mathcal{X}_{0}}\right) $ (see $\left( \ref%
{4.1}\right) $),%
\begin{equation}
\left\vert W\left( 0,r_{0}\right) \right\vert \leq \frac{\left\Vert
W_{0}\right\Vert _{\mathcal{X}_{0}}}{\left( 1+r_{0}\right) ^{2}}\leq \frac{%
4\left\Vert W_{0}\right\Vert _{\mathcal{X}_{0}}}{\left( 1+r_{1}+u_{1}\right)
^{2}l^{2}\left( x\right) }.  \tag{4.43}  \label{4.43}
\end{equation}%
Considering\ $\left( \ref{4.10}\right) $,\ $\left( \ref{4.30}\right) $,\ $%
\left( \ref{4.41}\right) $, and\ $\left( \ref{4.43}\right) $,\ we finally
arrive at the following estimate for the solution $w=\mathcal{K}\left(
W\right) $ of the initial value problem $\left( \ref{4.6}\right) $%
\begin{equation}
\begin{array}{l}
\left\vert w\left( u_{1},r_{1}\right) \right\vert \leq \exp \left(
\int_{0}^{u_{1}}\left\vert \left[ N\right] _{\gamma }\right\vert dv\right)
\left\vert W\left( 0,r_{0}\right) \right\vert +\int_{0}^{u_{1}}\left\{ \exp
\left( \int_{u}^{u_{1}}\left\vert \left[ N\right] _{\gamma }\right\vert
dv\right) \right\} \left\vert \left[ f\right] _{\gamma }\right\vert du \\ 
\text{ \ \ \ \ \ \ \ \ \ \ \ \ \ \ \ }\leq \frac{4\left\Vert
W_{0}\right\Vert _{\mathcal{X}_{0}}}{\left( 1+r_{1}+u_{1}\right)
^{2}l^{2}\left( x\right) }\exp \left( K_{1}\left(
x^{2}+x^{4}+x^{2p+2}\right) \right) \\ 
\text{ \ \ \ \ \ \ \ \ \ \ \ \ \ \ \ }+\exp \left( K_{1}\left(
x^{2}+x^{4}+x^{2p+2}\right) \right) \frac{\left( 12+56K_{0}\right) \left(
x^{3}+x^{5}+x^{2p+1}+x^{2p+3}\right) }{48\left( 1+u_{1}+r_{1}\right)
^{2}l^{2}\left( x\right) } \\ 
\text{ \ \ \ \ \ \ \ \ \ \ \ \ \ \ \ }\leq \left[ \frac{K_{2}\left(
x^{3}+x^{5}+x^{2p+1}+x^{2p+3}+\left\Vert W_{0}\right\Vert _{\mathcal{X}%
_{0}}\right) }{\left( 1+r_{1}+u_{1}\right) ^{2}l^{2}\left( x\right) }\right]
\exp \left( K_{1}\left( x^{2}+x^{4}+x^{2p+2}\right) \right) ,%
\end{array}
\tag{4.44}  \label{4.44}
\end{equation}%
where 
\begin{equation*}
K_{2}=\frac{204+56K_{0}}{48}.
\end{equation*}%
$\left( \ref{4.44}\right) $ yields%
\begin{equation}
\underset{r,u\geq 0}{\sup }\left[ \left( 1+u+r\right) ^{2}\left\vert w\left(
u,r\right) \right\vert \right] \leq \frac{K_{2}\left(
x^{3}+x^{5}+x^{2p+1}+x^{2p+3}+\left\Vert W_{0}\right\Vert _{\mathcal{X}%
_{0}}\right) }{l^{2}\left( x\right) }\exp \left( K_{1}\left(
x^{2}+x^{4}+x^{2p+2}\right) \right) .  \tag{4.45}  \label{4.45}
\end{equation}%
We also have to estimate $\underset{r,u\geq 0}{\sup }\left[ \left(
1+u+r\right) ^{3}\left\vert w^{\prime }\left( u,r\right) \right\vert \right] 
$. Set 
\begin{equation*}
z\left( u,r\right) =w^{\prime }\left( u,r\right) ,\text{\quad with }z\left(
0,r_{0}\right) =W^{\prime }\left( 0,r_{0}\right) .
\end{equation*}%
Differentiation of $\left( \ref{4.6}\right) $ w.r.t. $r$ gives%
\begin{equation}
Dz=N_{1}z+B_{1}w+B_{2}\overline{W}+B_{3}\overline{W}^{\prime },  \tag{4.46}
\label{4.46}
\end{equation}%
where the matrix functions $N_{1}$, $B_{1}$, $B_{2}$, and $B_{3}$ are given
by%
\begin{equation}
\begin{array}{l}
N_{1}=\left[ \frac{\widetilde{g}^{\prime }}{2}+\frac{g-\widetilde{g}}{2r}-%
\frac{Q^{2}g}{4r^{3}}-\frac{rVg}{2}\right] I_{2}+\frac{a}{2}i\sigma _{2}=%
\frac{\widetilde{g}^{\prime }}{2}I_{2}+N, \\ 
B_{1}=\left[ \frac{\left( g-\widetilde{g}\right) ^{\prime }}{2r}-\frac{g-%
\widetilde{g}}{2r^{2}}-\frac{QQ^{\prime }g}{2r^{3}}+\frac{3Q^{2}g}{4r^{4}}-%
\frac{Q^{2}g^{\prime }}{4r^{3}}-\frac{Vg+r\left( \Phi ^{\dag }\Phi \right)
^{\prime }V^{\prime }g+rVg^{\prime }}{2}\right] I_{2}+\frac{a^{\prime }}{2}%
i\sigma _{2}, \\ 
B_{2}=\left[ \frac{QQ^{\prime }g}{2r^{3}}-\frac{3Q^{2}g}{4r^{4}}+\frac{%
Q^{2}g^{\prime }}{4r^{3}}-\frac{\left( g-\widetilde{g}\right) ^{\prime }}{2r}%
+\frac{\left( g-\widetilde{g}\right) }{2r^{2}}+\frac{Vg+r\left( \Phi ^{\dag
}\Phi \right) ^{\prime }V^{\prime }g+rVg^{\prime }}{2}-\frac{gV^{\prime }+r%
\left[ V^{\prime }g^{\prime }+\left( \Phi ^{\dag }\Phi \right) ^{\prime
}V^{\prime \prime }g\right] }{2}\right] I_{2} \\ 
\text{ \ \ \ \ \ \ \ \ }+\left( \frac{Q^{\prime }g+Qg^{\prime }}{4r}-\frac{Qg%
}{4r^{2}}\right) \sigma _{1}, \\ 
B_{3}=\left[ \frac{Q^{2}g}{4r^{3}}-\frac{1}{2r}\left( g-\widetilde{g}\right)
+\frac{r\left( V-V^{\prime }\right) g}{2}\right] I_{2}+\frac{Qg}{4r}\sigma
_{1}.%
\end{array}
\tag{4.47}  \label{4.47}
\end{equation}%
Using the characteristics defined above the solution of $\left( \ref{4.46}%
\right) $ reads%
\begin{equation}
z\left( u_{1},r_{1}\right) =\exp \left( \int_{0}^{u_{1}}\left[ N_{1}\right]
_{\gamma }dv\right) W^{\prime }\left( 0,r_{0}\right)
+\int_{0}^{u_{1}}\left\{ \exp \left( \int_{u}^{u_{1}}\left[ N_{1}\right]
_{\gamma }dv\right) \right\} \left[ f_{1}\right] _{\gamma }du,  \tag{4.48}
\label{4.48}
\end{equation}%
where 
\begin{equation}
f_{1}=B_{1}w+B_{2}\overline{W}+B_{3}\overline{W}^{\prime }.  \tag{4.49}
\label{4.49}
\end{equation}%
Using $\left( \ref{3.6}\right) $ we calculate 
\begin{equation}
\frac{1}{2}\widetilde{g}^{\prime }=\frac{g-\overline{g}}{2r}-\frac{Q^{2}g}{%
4r^{3}}-\frac{rVg}{2}+\frac{1}{4r^{2}}\int_{0}^{r}\left( \frac{Q^{2}g}{s^{2}}%
\right) ds+\frac{1}{2r^{2}}\int_{0}^{r}s^{2}Vgds.  \tag{4.50}  \label{4.50}
\end{equation}%
From $\left( \ref{4.16}\right) $, $\left( \ref{4.17}\right) $, $\left( \ref%
{4.18}\right) $, $\left( \ref{4.19}\right) $, $\left( \ref{4.34}\right) $,
and $\left( \ref{4.50}\right) $, we\ gain\ the estimate%
\begin{equation}
\begin{array}{l}
\left\vert \frac{1}{2}\widetilde{g}^{\prime }\right\vert \leq \frac{x^{2}r}{%
24\left( 1+u\right) ^{3}\left( 1+u+r\right) ^{3}}+\frac{x^{4}r}{16\left(
1+u\right) ^{4}\left( 1+u+r\right) ^{4}}+\frac{K_{0}rx^{2p+2}}{2\left(
1+u\right) ^{5}\left( 1+u+r\right) ^{5}} \\ 
\text{ \ \ \ \ \ \ \ \ \ \ \ \ }+\frac{x^{4}r}{48\left( 1+u\right)
^{5}\left( 1+u+r\right) ^{3}}+\frac{K_{0}x^{2p+2}r}{6\left( 1+u\right)
^{6}\left( 1+u+r\right) ^{4}} \\ 
\text{ \ \ \ \ \ \ \ \ \ \ \ \ }\leq \frac{6+32K_{0}}{48}\frac{\left(
x^{2}+x^{4}+x^{2p+2}\right) r}{\left( 1+u\right) ^{3}\left( 1+u+r\right) ^{3}%
} \\ 
\text{ \ \ \ \ \ \ \ \ \ \ \ \ }\leq \frac{\left( 6+32K_{0}\right) \left(
x^{2}+x^{4}+x^{2p+2}\right) }{48}.%
\end{array}
\tag{4.51}  \label{4.51}
\end{equation}%
From $\left( \ref{4.36}\right) $ and $\left( \ref{4.51}\right) $, since $%
N_{1}=\frac{1}{2}\widetilde{g}^{\prime }+N$, we gain%
\begin{equation}
\left\vert N_{1}\left( u,r\right) \right\vert \leq \frac{36+64K_{0}}{48}%
\left( x^{2}+x^{4}+x^{2p+2}\right) .  \tag{4.52}  \label{4.52}
\end{equation}%
We now handle $f_{1}$. From $\left( \ref{3.3}\right) $\ and the\ definition\
of\ $g$ in$\ \left( \ref{3.9}\right) $ we have 
\begin{equation}
\left\vert g^{\prime }\right\vert \leq \left\vert \frac{\left( h-\overline{h}%
\right) ^{2}+\left( k-\overline{k}\right) ^{2}}{r}\right\vert \leq \frac{%
\left\vert W-\overline{W}\right\vert ^{2}}{r}\leq \frac{x^{2}r}{4\left(
1+u\right) ^{2}\left( 1+u+r\right) ^{4}}.  \tag{4.53}  \label{4.53}
\end{equation}%
$\left( \ref{4.51}\right) $, and $\left( \ref{4.53}\right) $ give 
\begin{equation}
\left\vert \frac{\left( g-\widetilde{g}\right) ^{\prime }}{2r}\right\vert
\leq \frac{\left( 6+16K_{0}\right) \left( x^{2}+x^{4}+x^{2p+2}\right) }{%
24\left( 1+u\right) ^{2}\left( 1+u+r\right) ^{3}}.  \tag{4.54}  \label{4.54}
\end{equation}%
$\left( \ref{4.20}\right) $ implies 
\begin{equation}
\left\vert \frac{g\left( u,r\right) -\widetilde{g}\left( u,r\right) }{2r^{2}}%
\right\vert \leq \frac{\left( 3+8K_{0}\right) \left(
x^{2}+x^{4}+x^{2p+2}\right) }{48\left( 1+u\right) ^{5}\left( 1+u+r\right)
^{3}}.  \tag{4.55}  \label{4.55}
\end{equation}%
In view of $\left( \ref{3.5}\right) $ and the definition of $\left\Vert
W\right\Vert _{\mathcal{X}}=x$, we have%
\begin{equation}
\left\vert Q^{\prime }\left( u,r\right) \right\vert \leq r\left\vert W\left(
u,r\right) \right\vert \left\vert \overline{W}\left( u,r\right) \right\vert
\leq \frac{x^{2}r}{\left( 1+u\right) \left( 1+u+r\right) ^{3}}.  \tag{4.56}
\label{4.56}
\end{equation}%
Since $0<g\leq 1$, we deduce from $\left( \ref{4.17}\right) $ and $\left( %
\ref{4.56}\right) $ that%
\begin{equation}
\left\vert \frac{QQ^{\prime }g}{2r^{3}}\right\vert \leq \frac{x^{4}}{4\left(
1+u\right) ^{3}\left( 1+u+r\right) ^{5}},  \tag{4.57}  \label{4.57}
\end{equation}%
and%
\begin{equation}
\left\vert \frac{3Q^{2}g}{4r^{4}}\right\vert \leq \frac{3x^{4}}{16\left(
1+u\right) ^{4}\left( 1+u+r\right) ^{4}}.  \tag{4.58}  \label{4.58}
\end{equation}%
From $\left( \ref{4.17}\right) $ and $\left( \ref{4.53}\right) $\ we obtain%
\begin{equation}
\left\vert \frac{Q^{2}g^{\prime }}{4r^{3}}\right\vert \leq \frac{x^{6}r^{2}}{%
64\left( 1+u\right) ^{6}\left( 1+u+r\right) ^{8}}\leq \frac{x^{6}}{64\left(
1+u\right) ^{6}\left( 1+u+r\right) ^{6}}.  \tag{4.59}  \label{4.59}
\end{equation}%
It is easy to see from $\left( \ref{2.10}\right) $ and\ $\left( \ref{3.2}%
\right) $\ that%
\begin{equation}
\left\vert \left( \Phi ^{\dag }\Phi \right) ^{\prime }\right\vert
=\left\vert \frac{2}{r}\left[ \overline{h}\left( h-\overline{h}\right) +%
\overline{k}\left( k-\overline{k}\right) \right] \right\vert \leq \frac{2}{r}%
\left\vert \overline{W}\left( u,r\right) \right\vert \left\vert W\left(
u,r\right) -\overline{W}\left( u,r\right) \right\vert .  \tag{4.60}
\label{4.60}
\end{equation}%
From $\left( \ref{4.2}\right) $, $\left( \ref{4.12}\right) $, $\left( \ref%
{4.13}\right) $, and $\left( \ref{4.60}\right) $ we\ derive\ the\ following\
estimates%
\begin{equation}
\begin{array}{l}
\left\vert \frac{Vg+r\left( \Phi ^{\dag }\Phi \right) ^{\prime }V^{\prime
}g+rVg^{\prime }}{2}\right\vert \\ 
\leq \frac{K_{0}\left\vert \overline{W}\left( u,r\right) \right\vert ^{2p+2}%
}{2}+\frac{K_{0}r\left\vert \overline{W}\left( u,r\right) \right\vert ^{2p}}{%
2}\frac{2}{r}\left\vert \overline{W}\left( u,r\right) \right\vert \left\vert
W\left( u,r\right) -\overline{W}\left( u,r\right) \right\vert +\frac{%
K_{0}r\left\vert \overline{W}\left( u,r\right) \right\vert ^{2p+2}}{2}\frac{%
x^{2}r}{2\left( 1+u\right) ^{2}\left( 1+u+r\right) ^{4}} \\ 
\leq \frac{K_{0}x^{2p+2}}{2\left( 1+u\right) ^{2p+2}\left( 1+u+r\right)
^{2p+2}}+\frac{K_{0}rx^{2p}}{2\left( 1+u\right) ^{2p}\left( 1+u+r\right)
^{2p}}\frac{2}{r}\frac{x}{\left( 1+u\right) \left( 1+u+r\right) }\frac{xr}{%
2\left( 1+u\right) \left( 1+u+r\right) ^{2}}+\frac{K_{0}r^{2}x^{2p+4}}{%
8\left( 1+u\right) ^{2p+4}\left( 1+u+r\right) ^{2p+6}} \\ 
\leq \frac{K_{0}x^{2p+2}}{2\left( 1+u\right) ^{2p+2}\left( 1+u+r\right)
^{2p+2}}+\frac{K_{0}x^{2p+2}}{2\left( 1+u\right) ^{2p+2}\left( 1+u+r\right)
^{2p+2}}+\frac{K_{0}x^{2p+4}}{8\left( 1+u\right) ^{2p+4}\left( 1+u+r\right)
^{2p+4}} \\ 
\leq \frac{K_{0}x^{2p+2}}{2\left( 1+u\right) ^{5}\left( 1+u+r\right) ^{5}}+%
\frac{K_{0}x^{2p+2}}{2\left( 1+u\right) ^{5}\left( 1+u+r\right) ^{5}}+\frac{%
K_{0}x^{2p+4}}{8\left( 1+u\right) ^{7}\left( 1+u+r\right) ^{7}}.%
\end{array}
\tag{4.61}  \label{4.61}
\end{equation}%
From $\left( \ref{3.4}\right) $ and $\left( \ref{4.17}\right) $\ we have%
\begin{equation}
\left\vert \frac{a^{\prime }}{2}\right\vert =\left\vert \frac{Qg}{2r^{2}}%
\right\vert \leq \frac{x^{2}}{4\left( 1+u\right) ^{2}\left( 1+u+r\right) ^{2}%
}.  \tag{4.62}  \label{4.62}
\end{equation}%
Inserting $\left( \ref{4.54}\right) $, $\left( \ref{4.55}\right) $, $\left( %
\ref{4.57}\right) $, $\left( \ref{4.58}\right) $, $\left( \ref{4.59}\right) $%
, $\left( \ref{4.61}\right) $, and$\ \left( \ref{4.62}\right) $ into the\
definition\ of\ $B_{1}$\ in $\left( \ref{4.47}\right) $ we obtain%
\begin{equation}
\begin{array}{l}
\left\vert B_{1}\right\vert \leq \frac{\left( 6+16K_{0}\right) \left(
x^{2}+x^{4}+x^{2p+2}\right) }{24\left( 1+u\right) ^{2}\left( 1+u+r\right)
^{3}}+\frac{\left( 3+8K_{0}\right) \left( x^{2}+x^{4}+x^{2p+2}\right) }{%
48\left( 1+u\right) ^{5}\left( 1+u+r\right) ^{3}}+\frac{x^{4}}{4\left(
1+u\right) ^{3}\left( 1+u+r\right) ^{5}} \\ 
\text{ \ \ \ \ \ \ \ \ }+\frac{3x^{4}}{16\left( 1+u\right) ^{4}\left(
1+u+r\right) ^{4}}+\frac{x^{6}}{64\left( 1+u\right) ^{6}\left( 1+u+r\right)
^{6}}+\frac{K_{0}x^{2p+2}}{2\left( 1+u\right) ^{5}\left( 1+u+r\right) ^{5}}
\\ 
\text{ \ \ \ \ \ \ \ \ }+\frac{K_{0}x^{2p+2}}{2\left( 1+u\right) ^{5}\left(
1+u+r\right) ^{5}}+\frac{K_{0}x^{2p+4}}{8\left( 1+u\right) ^{7}\left(
1+u+r\right) ^{7}}+\frac{x^{2}}{4\left( 1+u\right) ^{2}\left( 1+u+r\right)
^{2}} \\ 
\text{ \ \ \ \ \ \ \ \ }\leq \frac{K_{3}\left(
x^{2}+x^{4}+x^{6}+x^{2p}+x^{2p+2}+x^{2p+4}\right) }{\left( 1+u\right)
^{2}\left( 1+u+r\right) ^{2}},%
\end{array}
\tag{4.63}  \label{4.63}
\end{equation}%
where $K_{3}>0$ is\ an\ affine\ non-decreasing\ function\ of\ $K_{0}$. $%
\left( \ref{4.63}\right) $ yields%
\begin{equation}
\begin{array}{l}
\left\vert B_{1}w\left( u,r\right) \right\vert \leq \frac{K_{3}\left(
x^{2}+x^{4}+x^{2p}+x^{2p+2}+x^{2p+4}\right) }{\left( 1+u\right) ^{2}\left(
1+u+r\right) ^{2}}\left\vert w\left( u,r\right) \right\vert \\ 
\text{ \ \ \ \ \ \ \ \ \ \ \ \ \ \ \ \ \ }\leq \frac{K_{3}\left(
x^{2}+x^{4}+x^{6}+x^{2p}+x^{2p+2}+x^{2p+4}\right) }{\left( 1+u\right)
^{2}\left( 1+u+r\right) ^{4}}\underset{u,r\geq 0}{\sup }\left[ \left(
1+u+r\right) ^{2}\left\vert w\left( u,r\right) \right\vert \right] .%
\end{array}
\tag{4.64}  \label{4.64}
\end{equation}%
We now estimate $\left\vert B_{2}\overline{W}\left( u,r\right) \right\vert $%
. From $\left( \ref{4.2}\right) $, $\left( \ref{4.12}\right) $, $\left( \ref%
{4.13}\right) $, $\left( \ref{4.53}\right) $, and$\ \left( \ref{4.60}\right) 
$\ we get%
\begin{equation}
\begin{array}{l}
\left\vert \frac{gV^{\prime }+r\left[ V^{\prime }g^{\prime }+\left( \Phi
^{\dag }\Phi \right) ^{\prime }V^{\prime \prime }g\right] }{2}\right\vert
\leq \left\vert \frac{gV^{\prime }}{2}\right\vert +\left\vert \frac{%
rV^{\prime }g^{\prime }}{2}\right\vert +\left\vert \frac{r\left( \Phi ^{\dag
}\Phi \right) ^{\prime }V^{\prime \prime }g}{2}\right\vert \\ 
\text{ \ \ \ \ \ \ \ \ \ \ \ \ \ \ \ \ \ \ \ \ \ \ \ \ \ \ \ \ \ \ \ \ \ \ }%
\leq \left\vert \frac{gV^{\prime }}{2}\right\vert +\left\vert \frac{%
rV^{\prime }g^{\prime }}{2}\right\vert +\left\vert \frac{r\left( \Phi ^{\dag
}\Phi \right) ^{\prime }V^{\prime \prime }g}{2}\right\vert \\ 
\text{ \ \ \ \ \ \ \ \ \ \ \ \ \ \ \ \ \ \ \ \ \ \ \ \ \ \ \ \ \ \ \ \ \ \ }%
\leq \frac{K_{0}\left\vert \overline{W}\left( u,r\right) \right\vert ^{2p}}{2%
}+\frac{K_{0}r\left\vert \overline{W}\left( u,r\right) \right\vert ^{2p}}{2}%
\left\vert g^{\prime }\right\vert \\ 
\text{ \ \ \ \ \ \ \ \ \ \ \ \ \ \ \ \ \ \ \ \ \ \ \ \ \ \ \ \ \ \ \ \ \ \ }+%
\frac{K_{0}r\left\vert \overline{W}\left( u,r\right) \right\vert ^{2p-2}}{2}%
\frac{2}{r}\left\vert \overline{W}\left( u,r\right) \right\vert \left\vert
W\left( u,r\right) -\overline{W}\left( u,r\right) \right\vert \\ 
\text{ \ \ \ \ \ \ \ \ \ \ \ \ \ \ \ \ \ \ \ \ \ \ \ \ \ \ \ \ \ \ \ \ \ \ }%
\leq \frac{K_{0}x^{2p}}{2\left( 1+u\right) ^{2p}\left( 1+u+r\right) ^{2p}}+%
\frac{K_{0}rx^{2p}}{2\left( 1+u\right) ^{2p}\left( 1+u+r\right) ^{2p}}\frac{%
x^{2}r}{4\left( 1+u\right) ^{2}\left( 1+u+r\right) ^{4}} \\ 
\text{ \ \ \ \ \ \ \ \ \ \ \ \ \ \ \ \ \ \ \ \ \ \ \ \ \ \ \ \ \ \ \ \ \ \ }+%
\frac{K_{0}rx^{2p-2}}{2\left( 1+u\right) ^{2p-2}\left( 1+u+r\right) ^{2p-2}}%
\frac{2}{r}\frac{x}{\left( 1+u\right) \left( 1+u+r\right) }\frac{xr}{2\left(
1+u\right) \left( 1+u+r\right) ^{2}} \\ 
\text{ \ \ \ \ \ \ \ \ \ \ \ \ \ \ \ \ \ \ \ \ \ \ \ \ \ \ \ \ \ \ \ \ \ \ }%
\leq \frac{K_{0}x^{2p}}{2\left( 1+u\right) ^{3}\left( 1+u+r\right) ^{3}}+%
\frac{K_{0}x^{2p+2}}{8\left( 1+u\right) ^{5}\left( 1+u+r\right) ^{5}}+\frac{%
K_{0}x^{2p}}{2\left( 1+u\right) ^{3}\left( 1+u+r\right) ^{3}} \\ 
\text{ \ \ \ \ \ \ \ \ \ \ \ \ \ \ \ \ \ \ \ \ \ \ \ \ \ \ \ \ \ \ \ \ \ \ }%
\leq \frac{9K_{0}\left( x^{2p}+x^{2p+2}\right) }{8\left( 1+u\right)
^{3}\left( 1+u+r\right) ^{3}}.%
\end{array}
\tag{4.65}  \label{4.65}
\end{equation}%
$\left( \ref{4.17}\right) $, $\left( \ref{4.53}\right) $, and$\ \left( \ref%
{4.56}\right) $ give%
\begin{equation}
\begin{array}{l}
\left\vert \frac{Q^{\prime }g+Qg^{\prime }}{4r}-\frac{Qg}{4r^{2}}\right\vert
\leq \left\vert \frac{Q^{\prime }g}{4r}\right\vert +\left\vert \frac{%
Qg^{\prime }}{4r}\right\vert +\left\vert \frac{Qg}{4r^{2}}\right\vert \\ 
\text{ \ \ \ \ \ \ \ \ \ \ \ \ \ \ \ \ \ \ \ \ \ \ \ \ \ }\leq \frac{x^{2}}{%
4\left( 1+u\right) \left( 1+u+r\right) ^{3}}+\frac{x^{4}}{32\left(
1+u\right) ^{4}\left( 1+u+r\right) ^{4}}+\frac{x^{2}}{8\left( 1+u\right)
^{2}\left( 1+u+r\right) ^{2}} \\ 
\text{ \ \ \ \ \ \ \ \ \ \ \ \ \ \ \ \ \ \ \ \ \ \ \ \ \ }\leq \frac{%
13\left( x^{2}+x^{4}\right) }{32\left( 1+u\right) \left( 1+u+r\right) ^{2}}.%
\end{array}
\tag{4.66}  \label{4.66}
\end{equation}%
Inserting $\left( \ref{4.54}\right) $, $\left( \ref{4.55}\right) $, $\left( %
\ref{4.57}\right) $, $\left( \ref{4.58}\right) $, $\left( \ref{4.59}\right) $%
, $\left( \ref{4.61}\right) $, $\left( \ref{4.65}\right) $\ and$\ \left( \ref%
{4.66}\right) $ into the\ definition\ of\ $B_{2}$\ in $\left( \ref{4.47}%
\right) $ we obtain%
\begin{equation}
\left\vert B_{2}\right\vert \leq \frac{K_{4}\left(
x^{2}+x^{4}+x^{6}+x^{2p}+x^{2p+2}+x^{2p+4}\right) }{\left( 1+u\right) \left(
1+u+r\right) ^{2}},  \tag{4.67}  \label{4.67}
\end{equation}%
where $K_{4}>0$ is\ an\ affine\ non-decreasing\ function\ of\ $K_{0}$. $%
\left( \ref{4.12}\right) $ and $\left( \ref{4.67}\right) $ give%
\begin{equation}
\left\vert B_{2}\overline{W}\right\vert \leq \frac{K_{4}\left(
x^{3}+x^{5}+x^{7}+x^{2p+1}+x^{2p+3}+x^{2p+5}\right) }{\left( 1+u\right)
^{2}\left( 1+u+r\right) ^{3}}.  \tag{4.68}  \label{4.68}
\end{equation}%
We now handle $\left\vert B_{3}\overline{W}^{\prime }\left( u,r\right)
\right\vert $. From $\left( \ref{4.2}\right) $, $\left( \ref{4.17}\right) $, 
$\left( \ref{4.20}\right) $\ and the\ definition\ of $B_{3}$\ in $\left( \ref%
{4.47}\right) $\ we have%
\begin{equation}
\begin{array}{l}
\left\vert B_{3}\right\vert \leq \left\vert \frac{Q^{2}g}{4r^{3}}\right\vert
+\left\vert \frac{1}{2r}\left( g-\widetilde{g}\right) \right\vert
+\left\vert \frac{r\left( V-V^{\prime }\right) g}{2}\right\vert +\left\vert 
\frac{Qg}{4r}\right\vert \\ 
\text{ \ \ \ \ \ \ }\leq \frac{x^{4}}{16\left( 1+u\right) ^{4}\left(
1+u+r\right) ^{3}}+\frac{\left( 3+8K_{0}\right) \left(
x^{2}+x^{4}+x^{2p+2}\right) }{48\left( 1+u\right) ^{5}\left( 1+u+r\right)
^{2}} \\ 
\text{ \ \ \ \ \ \ }+\frac{K_{0}}{2}\left[ \frac{x^{2p+2}}{\left( 1+u\right)
^{5}\left( 1+u+r\right) ^{4}}+\frac{x^{2p}}{\left( 1+u\right) ^{3}\left(
1+u+r\right) ^{2}}\right] +\frac{x^{2}}{8\left( 1+u\right) ^{2}\left(
1+u+r\right) } \\ 
\text{ \ \ \ \ \ \ }\leq \frac{\left( 3+14K_{0}\right) \left(
x^{2}+x^{4}+x^{2p}+x^{2p+2}\right) }{12\left( 1+u\right) ^{2}\left(
1+u+r\right) }.%
\end{array}
\tag{4.69}  \label{4.69}
\end{equation}%
From\ $\left( \ref{4.13}\right) $ and $\left( \ref{4.69}\right) $\ we\ gain%
\begin{equation}
\begin{array}{l}
\left\vert B_{3}\overline{W}^{\prime }\left( u,r\right) \right\vert \leq 
\frac{\left( 3+14K_{0}\right) \left( x^{2}+x^{4}+x^{2p}+x^{2p+2}\right) }{12}%
\left\vert \frac{W-\overline{W}}{r}\right\vert \\ 
\text{ \ \ \ \ \ \ \ \ \ \ \ \ \ \ \ \ \ \ \ \ }\leq \frac{\left(
3+14K_{0}\right) \left( x^{3}+x^{5}+x^{2p+1}+x^{2p+3}\right) }{24\left(
1+u\right) ^{3}\left( 1+u+r\right) ^{3}}.%
\end{array}
\tag{4.70}  \label{4.70}
\end{equation}%
Inserting $\left( \ref{4.64}\right) $, $\left( \ref{4.68}\right) $,\ and$\
\left( \ref{4.70}\right) $ into the\ definition\ of\ $f_{1}$\ in $\left( \ref%
{4.49}\right) $, and\ using\ $\left( \ref{4.45}\right) $, we gain%
\begin{equation}
\begin{array}{l}
\left\vert f_{1}\left( u,r\right) \right\vert \leq \frac{K_{3}\left(
x^{2}+x^{4}+x^{6}+x^{2p}+x^{2p+2}+x^{2p+4}\right) }{\left( 1+u\right)
^{2}\left( 1+u+r\right) ^{4}}\underset{u,r\geq 0}{\sup }\left[ \left(
1+u+r\right) ^{2}\left\vert w\left( u,r\right) \right\vert \right] \\ 
\text{ \ \ \ \ \ \ \ \ \ \ \ \ \ \ }+\frac{K_{4}\left(
x^{3}+x^{5}+x^{7}+x^{2p+1}+x^{2p+3}+x^{2p+5}\right) }{\left( 1+u\right)
^{2}\left( 1+u+r\right) ^{3}}+\frac{\left( 3+14K_{0}\right) \left(
x^{3}+x^{5}+x^{2p+1}+x^{2p+3}\right) }{24\left( 1+u\right) ^{3}\left(
1+u+r\right) ^{3}} \\ 
\text{ \ \ \ \ \ \ \ \ \ \ \ \ \ \ }\leq \frac{K_{3}\left(
x^{2}+x^{4}+x^{6}+x^{2p}+x^{2p+2}+x^{2p+4}\right) }{\left( 1+u\right)
^{2}\left( 1+u+r\right) ^{4}}\frac{K_{2}\left(
x^{3}+x^{5}+x^{2p+1}+x^{2p+3}+\left\Vert W_{0}\right\Vert _{\mathcal{X}%
_{0}}\right) }{l^{2}\left( x\right) }\exp \left( K_{1}\left(
x^{2}+x^{4}+x^{2p+2}\right) \right) \\ 
\text{ \ \ \ \ \ \ \ \ \ \ \ \ \ \ }+\frac{K_{5}\left(
x^{3}+x^{5}+x^{7}+x^{2p+1}+x^{2p+3}+x^{2p+5}\right) }{\left( 1+u\right)
^{2}\left( 1+u+r\right) ^{3}},%
\end{array}
\tag{4.71}  \label{4.71}
\end{equation}%
where $K_{5}>0$ is\ an\ affine\ non-decreasing\ function\ of\ $K_{0}$. Using
similar tools as in $\left( \ref{4.30}\right) $ we have%
\begin{equation}
\begin{array}{l}
\int_{0}^{u_{1}}\left\vert \left[ f_{1}\right] _{\gamma }\right\vert
du=\int_{0}^{u_{1}}\left\vert f_{1}\left( u_{1},r_{1}\right) \right\vert du
\\ 
\text{ \ \ \ \ \ \ \ \ \ \ \ \ \ \ \ \ \ \ \ \ }\leq \frac{8K_{2}\left(
x^{3}+x^{5}+x^{2p+1}+x^{2p+3}+\left\Vert W_{0}\right\Vert _{\mathcal{X}%
_{0}}\right) K_{3}\left( x^{2}+x^{4}+x^{6}+x^{2p}+x^{2p+2}+x^{2p+4}\right)
\exp \left( K_{1}\left( x^{2}+x^{4}+x^{2p+2}\right) \right) }{l^{5}\left(
x\right) \left( 1+u_{1}+r_{1}\right) ^{3}} \\ 
\text{ \ \ \ \ \ \ \ \ \ \ \ \ \ \ \ \ \ \ \ \ }+\frac{8K_{5}\left(
x^{3}+x^{5}+x^{7}+x^{2p+1}+x^{2p+3}+x^{2p+5}\right) }{l^{3}\left( x\right)
\left( 1+u_{1}+r_{1}\right) ^{3}}.%
\end{array}
\tag{4.72}  \label{4.72}
\end{equation}%
Since $N_{1}=\frac{1}{2}\widetilde{g}^{\prime }+N$, it follows from $\left( %
\ref{4.41}\right) $ and $\left( \ref{4.51}\right) $\ that 
\begin{equation}
\left\vert \int_{0}^{u_{1}}\left[ N_{1}\right] _{\gamma }dv\right\vert \leq
K_{6}\left( x^{2}+x^{4}+x^{2p+2}\right) ,  \tag{4.73}  \label{4.73}
\end{equation}%
where $K_{6}>0$ is\ an\ affine\ non-decreasing\ function\ of\ $K_{0}$. By
analogy\ with\ $\left( \ref{4.43}\right) $ (see $\left( \ref{4.1}\right) $
for the definition of the Banach space $\left( \mathcal{X}_{0},\left\Vert
.\right\Vert _{\mathcal{X}_{0}}\right) $), it\ holds\ that%
\begin{equation}
\left\vert W^{\prime }\left( 0,r_{0}\right) \right\vert \leq \frac{%
\left\Vert W_{0}\right\Vert _{\mathcal{X}_{0}}}{\left( 1+r_{0}\right) ^{3}}%
\leq \frac{8\left\Vert W_{0}\right\Vert _{\mathcal{X}_{0}}}{\left(
1+r_{1}+u_{1}\right) ^{3}l^{3}\left( x\right) }.  \tag{4.74}  \label{4.74}
\end{equation}%
Considering $\left( \ref{4.72}\right) $,\ $\left( \ref{4.73}\right) $, and\ $%
\left( \ref{4.74}\right) $ we finally arrive at the following estimate for
the solution $z$ of the initial value problem $\left( \ref{4.46}\right) $%
\begin{equation}
\begin{array}{l}
\left\vert z\left( u_{1},r_{1}\right) \right\vert \leq \exp \left(
\int_{0}^{u_{1}}\left\vert \left[ N_{1}\right] _{\gamma }\right\vert
dv\right) \left\vert W^{\prime }\left( 0,r_{0}\right) \right\vert
+\int_{0}^{u_{1}}\left\{ \exp \left( \int_{u}^{u_{1}}\left\vert \left[ N_{1}%
\right] _{\gamma }\right\vert dv\right) \right\} \left\vert \left[ f_{1}%
\right] _{\gamma }\right\vert du \\ 
\leq \left[ \frac{8\left\Vert W_{0}\right\Vert _{\mathcal{X}_{0}}}{\left(
1+r_{1}+u_{1}\right) ^{3}l^{3}\left( x\right) }\right] \exp \left[
K_{6}\left( x^{2}+x^{4}+x^{2p+2}\right) \right] \\ 
+\frac{8K_{2}\left( x^{3}+x^{5}+x^{2p+1}+x^{2p+3}+\left\Vert
W_{0}\right\Vert _{\mathcal{X}_{0}}\right) K_{3}\left(
x^{2}+x^{4}+x^{6}+x^{2p}+x^{2p+2}+x^{2p+4}\right) \exp \left( K_{1}\left(
x^{2}+x^{4}+x^{2p+2}\right) \right) \exp \left[ K_{6}\left(
x^{2}+x^{4}+x^{2p+2}\right) \right] }{l^{5}\left( x\right) \left(
1+u_{1}+r_{1}\right) ^{3}} \\ 
+\frac{8K_{5}\left( x^{3}+x^{5}+x^{7}+x^{2p+1}+x^{2p+3}+x^{2p+5}\right) \exp %
\left[ K_{6}\left( x^{2}+x^{4}+x^{2p+2}\right) \right] }{l^{3}\left(
x\right) \left( 1+u_{1}+r_{1}\right) ^{3}} \\ 
\leq \left[ \frac{8\left\Vert W_{0}\right\Vert _{\mathcal{X}_{0}}}{\left(
1+r_{1}+u_{1}\right) ^{3}l^{3}\left( x\right) }+\frac{8K_{2}\left(
x^{3}+x^{5}+x^{2p+1}+x^{2p+3}+\left\Vert W_{0}\right\Vert _{\mathcal{X}%
_{0}}\right) K_{3}\left( x^{2}+x^{4}+x^{6}+x^{2p}+x^{2p+2}+x^{2p+4}\right) }{%
l^{5}\left( x\right) \left( 1+u_{1}+r_{1}\right) ^{3}}+\frac{8K_{5}\left(
x^{3}+x^{5}+x^{7}+x^{2p+1}+x^{2p+3}+x^{2p+5}\right) }{l^{3}\left( x\right)
\left( 1+u_{1}+r_{1}\right) ^{3}}\right] \\ 
\times \exp \left( K_{7}\left( x^{2}+x^{4}+x^{2p+2}\right) \right) ,%
\end{array}
\tag{4.75}  \label{4.75}
\end{equation}%
where $K_{7}>0$ is\ an\ affine\ non-decreasing\ function\ of\ $K_{0}$. From\ 
$\left( \ref{4.75}\right) $\ we\ deduce that%
\begin{equation}
\begin{array}{l}
\underset{u,r\geq 0}{\sup }\left[ \left( 1+r+u\right) ^{3}\left\vert z\left(
u,r\right) \right\vert \right] \\ 
\leq \left[ \frac{8\left\Vert W_{0}\right\Vert _{\mathcal{X}_{0}}}{%
l^{3}\left( x\right) }+\frac{8K_{2}\left(
x^{3}+x^{5}+x^{2p+1}+x^{2p+3}+\left\Vert W_{0}\right\Vert _{\mathcal{X}%
_{0}}\right) K_{3}\left( x^{2}+x^{4}+x^{6}+x^{2p}+x^{2p+2}+x^{2p+4}\right) }{%
l^{5}\left( x\right) }+\frac{8K_{5}\left(
x^{3}+x^{5}+x^{7}+x^{2p+1}+x^{2p+3}+x^{2p+5}\right) }{l^{3}\left( x\right) }%
\right] \\ 
\times \exp \left( K_{7}\left( x^{2}+x^{4}+x^{2p+2}\right) \right) .%
\end{array}
\tag{4.76}  \label{4.76}
\end{equation}%
Since $l\left( x\right) \in (0,1]$\ for all $x\in \lbrack 0,x_{0})$, it
follows from $\left( \ref{4.45}\right) $ and $\left( \ref{4.76}\right) $ that%
\begin{equation}
\begin{array}{l}
\left\Vert w\right\Vert _{\mathcal{X}}\leq \frac{K_{2}\left(
x^{3}+x^{5}+x^{2p+1}+x^{2p+3}+\left\Vert W_{0}\right\Vert _{\mathcal{X}%
_{0}}\right) }{l^{5}\left( x\right) }\exp \left( K_{9}\left(
x^{2}+x^{4}+x^{2p+2}\right) \right) \\ 
\text{ \ \ \ \ \ \ \ \ \ \ }+\frac{K_{8}\left(
x^{3}+x^{5}+x^{7}+x^{2p+1}+x^{2p+3}+x^{2p+5}+\left\Vert W_{0}\right\Vert _{%
\mathcal{X}_{0}}\right) \left(
x^{2}+x^{4}+x^{6}+x^{2p}+x^{2p+2}+x^{2p+4}\right) \exp \left( K_{9}\left(
x^{2}+x^{4}+x^{2p+2}\right) \right) }{l^{5}\left( x\right) },%
\end{array}
\tag{4.77}  \label{4.77}
\end{equation}%
where%
\begin{equation*}
K_{8}=\left( 8+8K_{4}+8K_{2}\right) K_{3},\quad K_{9}=K_{1}+K_{7}.
\end{equation*}%
Setting 
\begin{equation}
\Omega \left( x\right) =\frac{xl^{5}\left( x\right) \exp \left[ -K_{9}\left(
x^{2}+x^{4}+x^{2p+2}\right) \right] }{K_{2}+K_{8}\left(
x^{2}+x^{4}+x^{6}+x^{2p}+x^{2p+2}+x^{2p+4}\right) }-\left(
x^{3}+x^{5}+x^{7}+x^{2p+1}+x^{2p+3}+x^{2p+5}\right) ,  \tag{4.78}
\label{4.78}
\end{equation}%
one easily sees that%
\begin{equation*}
\Omega \left( 0\right) =0,\quad \Omega ^{\prime }\left( 0\right) >0,\quad 
\underset{x\longrightarrow +\infty }{\lim }\Omega \left( x\right) =-\infty .
\end{equation*}%
Hence there exists $x_{1}\in (0,x_{0})$ such that the function $\Omega $ is
strictly monotonically increasing on $[0,x_{1}]$. We then choose $x\in
(0,x_{1})$ and $\left\Vert W_{0}\right\Vert _{\mathcal{X}_{0}}<x$ to have $%
\left\Vert w\right\Vert _{\mathcal{X}}\leq x$. This shows that $\mathcal{K}$
is a mapping from $B_{\rho }$ into itself for $\rho \in (0,x_{1})$ and $%
\left\Vert W_{0}\right\Vert _{\mathcal{X}_{0}}<\rho $.

\subsubsection{The mapping $\mathcal{K}$ contracts in $\mathcal{Y}$}

Let $w_{1}$ and $w_{2}$ be two solutions of $\left( \ref{4.6}\right) $ with $%
w_{1}\left( 0,r\right) =w_{2}\left( 0,r\right) =W_{0}\left( r\right) ,$ $%
w_{j}=\mathcal{K}\left( W_{j}\right) ,$ $W_{j}\in \mathcal{X},$ $j=1,2$. We
use the following notations: $g_{j}=g\left( W_{j}\right) ,$ $Q_{j}=Q\left(
W_{j}\right) ,$ $V_{j}=V\left( W_{j}\right) ,\ z_{j}=z\left( W_{j}\right) $, 
$a_{j}=a\left( W_{j}\right) $. We assume 
\begin{equation*}
\max \left\{ \left\Vert W_{1}\right\Vert _{\mathcal{X}},\left\Vert
W_{2}\right\Vert _{\mathcal{X}}\right\} \leq x<x_{1},
\end{equation*}%
and set for convenience 
\begin{equation*}
w_{1}-w_{2}=\vartheta ,\quad W_{1}-W_{2}=\Theta ,\quad \left\Vert \Theta
\right\Vert _{\mathcal{Y}}=y,\quad D_{1}=\frac{\partial }{\partial u}-\frac{%
\widetilde{g_{1}}}{2}\frac{\partial }{\partial r}.
\end{equation*}%
Then we have the following non linear initial value problem with unknown $%
\vartheta $%
\begin{equation}
\begin{array}{l}
D_{1}\vartheta =N_{2}\vartheta +B_{4}\overline{\Theta }+B_{5}\left( w_{2}-%
\overline{W_{2}}\right) +B_{6}, \\ 
\vartheta \left( 0,r\right) =0,%
\end{array}
\tag{4.79}  \label{4.79}
\end{equation}%
where%
\begin{equation}
\begin{array}{l}
N_{2}=\left( \frac{g_{1}-\widetilde{g_{1}}}{2r}-\frac{Q_{1}^{2}g_{1}}{4r^{3}}%
-\frac{rV_{1}g_{1}}{2}\right) I_{2}+\frac{a_{1}}{2}i\sigma _{2} \\ 
B_{4}=\left( \frac{Q_{1}^{2}g_{1}}{4r^{3}}-\frac{g_{1}-\widetilde{g_{1}}}{2r}%
+\frac{rg_{1}\left( V_{1}-V_{1}^{\prime }\right) }{2}\right) I_{2}+\frac{%
Q_{1}g_{1}}{4r}\sigma _{1} \\ 
B_{5}=\frac{\left( g_{1}-\widetilde{g_{1}}\right) -\left( g_{2}-\widetilde{%
g_{2}}\right) }{2r}-\frac{\left( Q_{1}^{2}g_{1}-Q_{2}^{2}g_{2}\right) }{%
4r^{3}}-\frac{r\left( V_{1}g_{1}-V_{2}g_{2}\right) }{2} \\ 
B_{6}=-\frac{r\left( V_{1}^{\prime }g_{1}-V_{2}^{\prime }g_{2}\right) }{2}%
\overline{W_{2}}+\frac{Q_{1}g_{1}-Q_{2}g_{2}}{4r}\sigma _{1}\overline{W_{2}}+%
\frac{\widetilde{g_{1}}-\widetilde{g_{2}}}{2}\left( w_{2}\right) ^{\prime }+%
\frac{a_{1}-a_{2}}{2}i\sigma _{2}w_{2}.%
\end{array}
\tag{4.80}  \label{4.80}
\end{equation}%
Using the characteristic method as in the previous paragraph\ 4.2.2 we get%
\begin{equation}
\vartheta \left( u_{1},r_{1}\right) =\int_{0}^{u_{1}}\left\{ \exp \left(
\int_{u}^{u_{1}}\left[ N_{2}\left( u,r\right) \right] _{\gamma
_{1}}dv\right) \right\} \left[ f_{2}\right] _{\gamma _{1}}du,  \tag{4.81}
\label{4.81}
\end{equation}%
where%
\begin{equation}
f_{2}=B_{4}\overline{\Theta }+B_{5}\left( w_{2}-\overline{W_{2}}\right)
+B_{6},  \tag{4.82}  \label{4.82}
\end{equation}%
and $\gamma _{1}$ is the characteristic defined by%
\begin{equation}
\frac{dr}{du}=-\frac{1}{2}\widetilde{g_{1}}\left( u,\gamma _{1}\left(
u,r\right) \right) ,\quad r\left( 0\right) =r_{0}.  \tag{4.83}  \label{4.83}
\end{equation}%
We first estimate $N_{2}$ as in $\left( \ref{4.41}\right) $ to get%
\begin{equation}
\left\vert \int_{0}^{u_{1}}\left[ N_{2}\right] _{\gamma _{1}}dv\right\vert
\leq K_{1}\left( x^{2}+x^{4}+x^{2p+2}\right) .  \tag{4.84}  \label{4.84}
\end{equation}%
We now concentrate on estimating $f_{2}$. Similarly to $\left( \ref{4.27}%
\right) $ it is straightforward to have%
\begin{equation}
\left\vert B_{4}\overline{\Theta }\right\vert \leq \frac{\left(
12+56K_{0}\right) \left( x^{2}+x^{4}+x^{2p}+x^{2p+2}\right) y}{48\left(
1+u\right) ^{3}\left( 1+u+r\right) ^{2}}.  \tag{4.85}  \label{4.85}
\end{equation}%
Using the mean value theorem and similar calculation as in $\left( \ref{4.13}%
\right) $ we have%
\begin{equation}
\left\vert g_{1}-g_{2}\right\vert \leq \int_{r}^{+\infty }\frac{\left(
\left\vert \Theta \right\vert +\left\vert \overline{\Theta }\right\vert
\right) \left( \left\vert W_{1}-\overline{W_{1}}\right\vert +\left\vert
W_{2}-\overline{W_{2}}\right\vert \right) }{s}ds\leq \frac{xy}{3\left(
1+u\right) ^{2}\left( 1+u+r\right) ^{2}},  \tag{4.86}  \label{4.86}
\end{equation}%
and%
\begin{equation}
\left\vert \overline{g_{1}}-\overline{g_{2}}\right\vert \leq \frac{1}{r}%
\int_{0}^{r}\frac{xy}{3\left( 1+u\right) ^{2}\left( 1+u+s\right) ^{2}}ds=%
\frac{xy}{3\left( 1+u\right) ^{3}\left( 1+u+r\right) }.  \tag{4.87}
\label{4.87}
\end{equation}%
In the same way, using the mean value theorem and similar calculation as in $%
\left( \ref{4.12}\right) $ we have 
\begin{equation}
\left\vert Q_{1}-Q_{2}\right\vert \leq \int_{0}^{r}s\left( \left\vert 
\overline{W_{1}}\right\vert \left\vert \Theta \right\vert +\left\vert
W_{1}\right\vert \left\vert \overline{\Theta }\right\vert \right) ds\leq 
\frac{xyr^{2}}{\left( 1+u\right) ^{2}\left( 1+u+r\right) ^{2}}.  \tag{4.88}
\label{4.88}
\end{equation}%
From $\left( \ref{4.17}\right) $, $\left( \ref{4.86}\right) $, $\left( \ref%
{4.87}\right) $, and $\left( \ref{4.88}\right) $\ we have 
\begin{equation*}
\begin{array}{l}
\left\vert Q_{1}^{2}g_{1}-Q_{2}^{2}g_{2}\right\vert =\left\vert
Q_{1}^{2}\left( g_{1}-g_{2}\right) +\left( Q_{1}-Q_{2}\right) \left(
Q_{1}+Q_{2}\right) g_{2}\right\vert \\ 
\text{ \ \ \ \ \ \ \ \ \ \ \ \ \ \ \ \ \ \ \ \ \ }\leq \left\vert
Q_{1}\right\vert ^{2}\left\vert g_{1}-g_{2}\right\vert +\left\vert
Q_{1}-Q_{2}\right\vert \left( \left\vert Q_{1}\right\vert +\left\vert
Q_{2}\right\vert \right) \\ 
\text{ \ \ \ \ \ \ \ \ \ \ \ \ \ \ \ \ \ \ \ \ \ }\leq \frac{x^{5}yr^{4}}{%
12\left( 1+u\right) ^{6}\left( 1+u+r\right) ^{6}}+\frac{x^{3}yr^{4}}{\left(
1+u\right) ^{4}\left( 1+u+r\right) ^{4}} \\ 
\text{ \ \ \ \ \ \ \ \ \ \ \ \ \ \ \ \ \ \ \ \ \ }\leq \frac{13\left(
x^{3}+x^{5}\right) yr^{4}}{12\left( 1+u\right) ^{4}\left( 1+u+r\right) ^{4}},%
\end{array}%
\end{equation*}%
which\ implies%
\begin{equation}
\left\vert \frac{\left( Q_{1}^{2}g_{1}-Q_{2}^{2}g_{2}\right) }{4r^{3}}%
\right\vert \leq \frac{13\left( x^{3}+x^{5}\right) yr}{48\left( 1+u\right)
^{4}\left( 1+u+r\right) ^{4}}.  \tag{4.89}  \label{4.89}
\end{equation}%
From $\left( \ref{3.6}\right) $ and\ the\ definition\ of $\widetilde{g}$\
in\ $\left( \ref{3.9}\right) $\ we have%
\begin{equation}
\left\vert \widetilde{g_{1}}-\widetilde{g_{2}}\right\vert \leq \left\vert 
\overline{g_{1}}-\overline{g_{2}}\right\vert +\frac{1}{2r}\int_{0}^{r}\frac{%
\left\vert Q_{1}^{2}g_{1}-Q_{2}^{2}g_{2}\right\vert }{s^{2}}ds+\frac{1}{r}%
\int_{0}^{r}s^{2}\left\vert V_{1}g_{1}-V_{2}g_{2}\right\vert ds.  \tag{4.90}
\label{4.90}
\end{equation}%
$\left( \ref{4.89}\right) $ yields%
\begin{equation}
\frac{1}{2r}\int_{0}^{r}\frac{\left\vert
Q_{1}^{2}g_{1}-Q_{2}^{2}g_{2}\right\vert }{s^{2}}ds\leq \frac{13\left(
x^{3}+x^{5}\right) y}{24r\left( 1+u\right) ^{4}}\int_{0}^{r}\frac{s^{2}}{%
\left( 1+u+s\right) ^{4}}ds=\frac{13\left( x^{3}+x^{5}\right) yr^{2}}{%
72\left( 1+u\right) ^{5}\left( 1+u+r\right) ^{3}}.  \tag{4.91}  \label{4.91}
\end{equation}%
Using the mean value theorem and assumption $\left( \ref{4.2}\right) $ we get%
\begin{equation}
\begin{array}{l}
\left\vert V_{1}-V_{2}\right\vert =\left\vert V\left( \left( \overline{h_{1}}%
\right) ^{2}+\left( \overline{k_{1}}\right) ^{2}\right) -V\left( \left( 
\overline{h_{2}}\right) ^{2}+\left( \overline{k_{2}}\right) ^{2}\right)
\right\vert \\ 
\leq \int_{0}^{1}\left\vert \left( \left( \overline{h_{2}}\right)
^{2}+\left( \overline{k_{2}}\right) ^{2}\right) -\left( \left( \overline{%
h_{1}}\right) ^{2}+\left( \overline{k_{1}}\right) ^{2}\right) \right\vert
\left\vert V^{\prime }\left[ \left( 1-t\right) \left( \left( \overline{h_{1}}%
\right) ^{2}+\left( \overline{k_{1}}\right) ^{2}\right) +t\left( \left( 
\overline{h_{2}}\right) ^{2}+\left( \overline{k_{2}}\right) ^{2}\right) %
\right] \right\vert dt \\ 
\leq K_{0}\left\vert \overline{\Theta }\right\vert \left( \left\vert 
\overline{W_{1}}\right\vert +\left\vert \overline{W_{2}}\right\vert \right)
\left( \left\vert \overline{W_{1}}\right\vert ^{2p}+\left\vert \overline{%
W_{2}}\right\vert ^{2p}\right) \\ 
\leq \frac{4K_{0}x^{2p+1}y}{\left( 1+u\right) ^{2p+1}\left( 1+u+r\right)
^{2p+1}} \\ 
\leq \frac{4K_{0}x^{2p+1}y}{\left( 1+u\right) ^{4}\left( 1+u+r\right) ^{4}}.%
\end{array}
\tag{4.92}  \label{4.92}
\end{equation}%
$\left( \ref{4.2}\right) $, $\left( \ref{4.86}\right) $, and $\left( \ref%
{4.92}\right) $ give%
\begin{equation}
\begin{array}{l}
\left\vert V_{1}g_{1}-V_{2}g_{2}\right\vert =\left\vert V_{1}\left(
g_{1}-g_{2}\right) +\left( V_{1}-V_{2}\right) g_{2}\right\vert \\ 
\text{ \ \ \ \ \ \ \ \ \ \ \ \ \ \ \ \ \ \ \ \ \ \ }\leq \left\vert
V_{1}\right\vert \left\vert g_{1}-g_{2}\right\vert +\left\vert
V_{1}-V_{2}\right\vert \\ 
\text{ \ \ \ \ \ \ \ \ \ \ \ \ \ \ \ \ \ \ \ \ \ \ }\leq \frac{K_{0}x^{2p+3}y%
}{3\left( 1+u\right) ^{7}\left( 1+u+r\right) ^{7}}+\frac{4K_{0}x^{2p+1}y}{%
\left( 1+u\right) ^{4}\left( 1+u+r\right) ^{4}} \\ 
\text{ \ \ \ \ \ \ \ \ \ \ \ \ \ \ \ \ \ \ \ \ \ \ }\leq \frac{\left(
1+12K_{0}\right) \left( x^{2p+1}+x^{2p+3}\right) y}{3\left( 1+u\right)
^{4}\left( 1+u+r\right) ^{4}}.%
\end{array}
\tag{4.93}  \label{4.93}
\end{equation}%
Thus%
\begin{equation}
\begin{array}{l}
\frac{1}{r}\int_{0}^{r}s^{2}\left\vert V_{1}g_{1}-V_{2}g_{2}\right\vert
ds\leq \frac{\left( 1+12K_{0}\right) \left( x^{2p+1}+x^{2p+3}\right) y}{%
3\left( 1+u\right) ^{4}r}\int_{0}^{r}\frac{s^{2}}{\left( 1+u+s\right) ^{4}}ds
\\ 
\text{ \ \ \ \ \ \ \ \ \ \ \ \ \ \ \ \ \ \ \ \ \ \ \ \ \ \ \ \ \ \ \ \ \ \ \
\ \ }=\frac{\left( 1+12K_{0}\right) \left( x^{2p+1}+x^{2p+3}\right) r^{2}y}{%
9\left( 1+u\right) ^{5}\left( 1+u+r\right) ^{3}}%
\end{array}
\tag{4.94}  \label{4.94}
\end{equation}%
It follows from $\left( \ref{4.87}\right) $, $\left( \ref{4.90}\right) $, $%
\left( \ref{4.91}\right) $, and $\left( \ref{4.94}\right) $ that%
\begin{equation}
\left\vert \widetilde{g_{1}}-\widetilde{g_{2}}\right\vert \leq \frac{\left(
15+32K_{0}\right) \left( x+x^{3}+x^{5}+x^{2p+1}+x^{2p+3}\right) y}{24\left(
1+u\right) ^{3}\left( 1+u+r\right) }.  \tag{4.95}  \label{4.95}
\end{equation}%
By analogy with $\left( \ref{4.16}\right) $ it holds that%
\begin{equation}
\begin{array}{l}
\left\vert g_{1}-g_{2}-\left( \overline{g_{1}}-\overline{g_{2}}\right)
\right\vert \leq \frac{1}{r}\int_{0}^{r}\left\vert \left( g_{1}-g_{2}\right)
\left( u,r\right) -\left( g_{1}-g_{2}\right) \left( u,r^{\prime }\right)
\right\vert dr^{\prime } \\ 
\text{ \ \ \ \ \ \ \ \ \ \ \ \ \ \ \ \ \ \ \ \ \ \ \ \ \ \ \ \ \ \ \ \ }\leq 
\frac{1}{r}\int_{0}^{r}\left[ \int_{r^{\prime }}^{r}\left\vert \frac{%
\partial \left( g_{1}-g_{2}\right) \left( u,s\right) }{\partial s}%
\right\vert ds\right] dr^{\prime } \\ 
\text{ \ \ \ \ \ \ \ \ \ \ \ \ \ \ \ \ \ \ \ \ \ \ \ \ \ \ \ \ \ \ \ \ }\leq 
\frac{1}{r}\int_{0}^{r}\left[ \int_{r^{\prime }}^{r}\frac{\left( \left\vert
\Theta \right\vert +\left\vert \overline{\Theta }\right\vert \right) \left(
\left\vert W_{1}-\overline{W_{1}}\right\vert +\left\vert W_{2}-\overline{%
W_{2}}\right\vert \right) }{s}ds\right] dr^{\prime } \\ 
\text{ \ \ \ \ \ \ \ \ \ \ \ \ \ \ \ \ \ \ \ \ \ \ \ \ \ \ \ \ \ \ \ \ }\leq 
\frac{1}{r}\int_{0}^{r}\left[ \int_{r^{\prime }}^{r}\frac{2xy}{\left(
1+u\right) ^{2}\left( 1+u+s\right) ^{3}}ds\right] dr^{\prime } \\ 
\text{ \ \ \ \ \ \ \ \ \ \ \ \ \ \ \ \ \ \ \ \ \ \ \ \ \ \ \ \ \ \ \ \ }=%
\frac{2xyr}{\left( 1+u\right) ^{3}\left( 1+u+r\right) ^{2}}.%
\end{array}
\tag{4.96}  \label{4.96}
\end{equation}%
Combination of $\left( \ref{4.91}\right) $, $\left( \ref{4.94}\right) $, and 
$\left( \ref{4.96}\right) $ yields%
\begin{equation}
\left\vert \frac{\left( g_{1}-\widetilde{g_{1}}\right) -\left( g_{2}-%
\widetilde{g_{2}}\right) }{2r}\right\vert \leq \frac{\left(
31+32K_{0}\right) \left( x+x^{3}+x^{5}+x^{2p+1}+x^{2p+3}\right) y}{24\left(
1+u\right) ^{3}\left( 1+u+r\right) ^{2}}.  \tag{4.97}  \label{4.97}
\end{equation}%
In view of $\left( \ref{4.89}\right) $, $\left( \ref{4.93}\right) $, $\left( %
\ref{4.97}\right) $, and\ the\ definition\ of $B_{5}$\ in\ $\left( \ref{4.80}%
\right) $ we gain%
\begin{equation}
\begin{array}{l}
\left\vert B_{5}\right\vert \leq \left\vert \frac{\left( g_{1}-\widetilde{%
g_{1}}\right) -\left( g_{2}-\widetilde{g_{2}}\right) }{2r}\right\vert
+\left\vert \frac{\left( Q_{1}^{2}g_{1}-Q_{2}^{2}g_{2}\right) }{4r^{3}}%
\right\vert +\left\vert \frac{r\left( V_{1}g_{1}-V_{2}g_{2}\right) }{2}%
\right\vert \\ 
\text{ \ \ \ \ }\leq \frac{\left( 31+32K_{0}\right) \left(
x+x^{3}+x^{5}+x^{2p+1}+x^{2p+3}\right) y}{24\left( 1+u\right) ^{3}\left(
1+u+r\right) ^{2}}+\frac{13\left( x^{3}+x^{5}\right) yr}{48\left( 1+u\right)
^{4}\left( 1+u+r\right) ^{4}}+\frac{\left( 1+12K_{0}\right) \left(
x^{2p+1}+x^{2p+3}\right) ry}{6\left( 1+u\right) ^{4}\left( 1+u+r\right) ^{4}}
\\ 
\text{ \ \ \ \ \ }\leq \frac{\left( 83+160K_{0}\right) \left(
x+x^{3}+x^{5}+x^{2p+1}+x^{2p+3}\right) y}{24\left( 1+u\right) ^{3}\left(
1+u+r\right) ^{2}}.%
\end{array}
\tag{4.98}  \label{4.98}
\end{equation}%
From $\left( \ref{4.12}\right) $ and\ $\left( \ref{4.98}\right) $ we obtain%
\begin{equation}
\begin{array}{l}
\left\vert B_{5}\left( w_{2}-\overline{W_{2}}\right) \right\vert \leq
\left\vert B_{5}\right\vert \left\vert w_{2}\right\vert +\left\vert
B_{5}\right\vert \left\vert \overline{W_{2}}\right\vert \\ 
\text{ \ \ \ \ \ \ \ \ \ \ \ \ \ \ \ \ \ \ \ \ \ \ }\leq \frac{\left(
83+160K_{0}\right) \left( x+x^{3}+x^{5}+x^{2p+1}+x^{2p+3}\right) y}{24\left(
1+u\right) ^{3}\left( 1+u+r\right) ^{4}}\underset{u,r\geq 0}{\sup }\left\{
\left( 1+u+r\right) ^{2}\left\vert w_{2}\right\vert \right\} \\ 
\text{ \ \ \ \ \ \ \ \ \ \ \ \ \ \ \ \ \ \ \ \ \ \ }+\frac{\left(
83+160K_{0}\right) \left( x^{2}+x^{4}+x^{6}+x^{2p+2}+x^{2p+4}\right) y}{%
24\left( 1+u\right) ^{4}\left( 1+u+r\right) ^{5}} \\ 
\text{ \ \ \ \ \ \ \ \ \ \ \ \ \ \ \ \ \ \ \ \ \ \ }\leq \frac{\left(
83+160K_{0}\right) \left( x^{2}+x^{4}+x^{6}+x^{2p+2}+x^{2p+4}\right) y}{%
24\left( 1+u\right) ^{3}\left( 1+u+r\right) ^{4}}+\frac{\left(
83+160K_{0}\right) \left( x^{2}+x^{4}+x^{6}+x^{2p+2}+x^{2p+4}\right) y}{%
24\left( 1+u\right) ^{4}\left( 1+u+r\right) ^{5}} \\ 
\text{ \ \ \ \ \ \ \ \ \ \ \ \ \ \ \ \ \ \ \ \ \ \ }\leq \frac{\left(
83+160K_{0}\right) \left( x^{2}+x^{4}+x^{6}+x^{2p+2}+x^{2p+4}\right) y}{%
12\left( 1+u\right) ^{3}\left( 1+u+r\right) ^{4}}%
\end{array}
\tag{4.99}  \label{4.99}
\end{equation}%
We now estimate $B_{6}$. Using once more the mean value theorem and
assumption $\left( \ref{4.2}\right) $ we gain%
\begin{equation}
\begin{array}{l}
\left\vert V_{1}^{\prime }-V_{2}^{\prime }\right\vert =\left\vert V^{\prime
}\left( \left( \overline{h_{1}}\right) ^{2}+\left( \overline{k_{1}}\right)
^{2}\right) -V^{\prime }\left( \left( \overline{h_{2}}\right) ^{2}+\left( 
\overline{k_{2}}\right) ^{2}\right) \right\vert \\ 
\leq \int_{0}^{1}\left\vert \left( \left( \overline{h_{2}}\right)
^{2}+\left( \overline{k_{2}}\right) ^{2}\right) -\left( \left( \overline{%
h_{1}}\right) ^{2}+\left( \overline{k_{1}}\right) ^{2}\right) \right\vert
\left\vert V^{\prime \prime }\left[ \left( 1-t\right) \left( \left( 
\overline{h_{1}}\right) ^{2}+\left( \overline{k_{1}}\right) ^{2}\right)
+t\left( \left( \overline{h_{2}}\right) ^{2}+\left( \overline{k_{2}}\right)
^{2}\right) \right] \right\vert dt \\ 
\leq K_{0}\left\vert \overline{\Theta }\right\vert \left( \left\vert 
\overline{W_{1}}\right\vert +\left\vert \overline{W_{2}}\right\vert \right)
\left( \left\vert \overline{W_{1}}\right\vert ^{2p-2}+\left\vert \overline{%
W_{2}}\right\vert ^{2p-2}\right) \\ 
\leq \frac{4K_{0}x^{2p-1}y}{\left( 1+u\right) ^{2p-1}\left( 1+u+r\right)
^{2p-1}} \\ 
\leq \frac{4K_{0}x^{2p-1}y}{\left( 1+u\right) ^{2}\left( 1+u+r\right) ^{2}}.%
\end{array}
\tag{4.100}  \label{4.100}
\end{equation}%
$\left( \ref{4.2}\right) $, $\left( \ref{4.86}\right) $, and $\left( \ref%
{4.100}\right) $%
\begin{equation}
\begin{array}{l}
\left\vert V_{1}^{\prime }g_{1}-V_{2}^{\prime }g_{2}\right\vert =\left\vert
V_{1}^{\prime }\left( g_{1}-g_{2}\right) +\left( V_{1}^{\prime
}-V_{2}^{\prime }\right) g_{2}\right\vert \\ 
\text{ \ \ \ \ \ \ \ \ \ \ \ \ \ \ \ \ \ \ \ \ \ }\leq \left\vert
V_{1}^{\prime }\right\vert \left\vert g_{1}-g_{2}\right\vert +\left\vert
V_{1}^{\prime }-V_{2}^{\prime }\right\vert \\ 
\text{ \ \ \ \ \ \ \ \ \ \ \ \ \ \ \ \ \ \ \ \ \ }\leq \frac{K_{0}x^{2p+1}y}{%
3\left( 1+u\right) ^{5}\left( 1+u+r\right) ^{5}}+\frac{4K_{0}x^{2p-1}y}{%
\left( 1+u\right) ^{2}\left( 1+u+r\right) ^{2}} \\ 
\text{ \ \ \ \ \ \ \ \ \ \ \ \ \ \ \ \ \ \ \ \ \ }\leq \frac{\left(
1+12K_{0}\right) \left( x^{2p-1}+x^{2p+1}\right) y}{3\left( 1+u\right)
^{2}\left( 1+u+r\right) ^{2}}.%
\end{array}
\tag{4.101}  \label{4.101}
\end{equation}%
$\left( \ref{4.12}\right) $, and $\left( \ref{4.101}\right) $ yield%
\begin{equation}
\left\vert \frac{r\left( V_{1}^{\prime }g_{1}-V_{2}^{\prime }g_{2}\right) }{2%
}\overline{W_{2}}\right\vert \leq \frac{\left( 1+12K_{0}\right) \left(
x^{2p}+x^{2p+2}\right) y}{6\left( 1+u\right) ^{3}\left( 1+u+r\right) ^{2}}. 
\tag{4.102}  \label{4.102}
\end{equation}%
From $\left( \ref{4.12}\right) $, $\left( \ref{4.17}\right) $, $\left( \ref%
{4.86}\right) $, and $\left( \ref{4.88}\right) $ we\ derive\ the\ following\
estimate%
\begin{equation}
\begin{array}{l}
\left\vert \frac{Q_{1}g_{1}-Q_{2}g_{2}}{4r}\sigma _{1}\overline{W_{2}}%
\right\vert \leq \left\vert \frac{Q_{1}g_{1}-Q_{2}g_{2}}{4r}\right\vert
\left\vert \sigma _{1}\overline{W_{2}}\right\vert \\ 
\text{ \ \ \ \ \ \ \ \ \ \ \ \ \ \ \ \ \ \ \ \ \ \ \ \ \ \ \ }\leq \left[
\left( \left\vert Q_{1}\left( g_{1}-g_{2}\right) \right\vert +\left\vert
\left( Q_{1}-Q_{2}\right) g_{2}\right\vert \right) \right] \frac{\left\vert
\sigma _{1}\overline{W_{2}}\right\vert }{4r} \\ 
\text{ \ \ \ \ \ \ \ \ \ \ \ \ \ \ \ \ \ \ \ \ \ \ \ \ \ \ \ }\leq \frac{%
x^{4}yr}{24\left( 1+u\right) ^{5}\left( 1+u+r\right) ^{5}}+\frac{x^{2}yr}{%
4\left( 1+u\right) ^{3}\left( 1+u+r\right) ^{3}} \\ 
\text{ \ \ \ \ \ \ \ \ \ \ \ \ \ \ \ \ \ \ \ \ \ \ \ \ \ \ \ }\leq \frac{%
7\left( x^{2}+x^{4}\right) y}{24\left( 1+u\right) ^{3}\left( 1+u+r\right)
^{2}}.%
\end{array}
\tag{4.103}  \label{4.103}
\end{equation}%
In view of $\left( \ref{4.95}\right) $ we have%
\begin{equation}
\begin{array}{l}
\left\vert \frac{\widetilde{g_{1}}-\widetilde{g_{2}}}{2}\left( w_{2}\right)
^{\prime }\right\vert \leq \frac{\left( 15+32K_{0}\right) \left(
x+x^{3}+x^{5}+x^{2p+1}+x^{2p+3}\right) y}{48\left( 1+u\right) ^{3}\left(
1+u+r\right) }\left\vert \left( w_{2}\right) ^{\prime }\right\vert \\ 
\text{ \ \ \ \ \ \ \ \ \ \ \ \ \ \ \ \ \ \ \ \ }\leq \frac{\left(
15+32K_{0}\right) \left( x+x^{3}+x^{5}+x^{2p+1}+x^{2p+3}\right) y}{48\left(
1+u\right) ^{3}\left( 1+u+r\right) ^{4}}\underset{u,r\geq 0}{\sup }\left\{
\left( 1+u+r\right) ^{3}\left\vert \left( w_{2}\right) ^{\prime }\right\vert
\right\} \\ 
\text{ \ \ \ \ \ \ \ \ \ \ \ \ \ \ \ \ \ \ \ \ }\leq \frac{\left(
15+32K_{0}\right) \left( x^{2}+x^{4}+x^{6}+x^{2p+2}+x^{2p+4}\right) y}{%
48\left( 1+u\right) ^{3}\left( 1+u+r\right) ^{4}}.%
\end{array}
\tag{4.104}  \label{4.104}
\end{equation}%
We use $\left( \ref{3.4}\right) $, $\left( \ref{4.86}\right) $, and $\left( %
\ref{4.88}\right) $\ to gain%
\begin{equation}
\begin{array}{l}
\left\vert a_{1}-a_{2}\right\vert =\left\vert \int_{0}^{r}\frac{%
Q_{1}g_{1}-Q_{2}g_{2}}{s^{2}}ds\right\vert \\ 
\text{ \ \ \ \ \ \ \ \ \ \ \ \ \ }\leq \int_{0}^{r}\frac{\left\vert
Q_{1}\left( g_{1}-g_{2}\right) \right\vert }{s^{2}}ds+\int_{0}^{r}\frac{%
\left\vert \left( Q_{1}-Q_{2}\right) g_{2}\right\vert }{s^{2}}ds \\ 
\text{ \ \ \ \ \ \ \ \ \ \ \ \ \ }\leq \frac{x^{3}y}{6\left( 1+u\right) ^{4}}%
\int_{0}^{r}\frac{1}{\left( 1+u+s\right) ^{4}}ds+\frac{xy}{\left( 1+u\right)
^{2}}\int_{0}^{r}\frac{1}{\left( 1+u+s\right) ^{2}}ds \\ 
\text{ \ \ \ \ \ \ \ \ \ \ \ \ \ }=\frac{x^{3}y\left[ 3r\left( 1+u\right)
^{2}+3r^{2}\left( 1+u\right) +r^{3}\right] }{18\left( 1+u\right) ^{7}\left(
1+u+r\right) ^{3}}+\frac{xyr}{\left( 1+u\right) ^{3}\left( 1+u+r\right) } \\ 
\text{ \ \ \ \ \ \ \ \ \ \ \ \ \ }\leq \frac{19\left( x+x^{3}\right) y}{%
18\left( 1+u\right) ^{3}}.%
\end{array}
\tag{4.105}  \label{4.105}
\end{equation}%
Since $\max \left\{ \left\Vert W_{1}\right\Vert _{\mathcal{X}},\left\Vert
W_{2}\right\Vert _{\mathcal{X}}\right\} \leq x<x_{1}$, $\left( \ref{4.105}%
\right) $\ implies%
\begin{equation}
\begin{array}{l}
\left\vert \frac{a_{1}-a_{2}}{2}i\sigma _{2}w_{2}\right\vert \leq \frac{%
19\left( x+x^{3}\right) y}{36\left( 1+u\right) ^{3}}\left\vert
w_{2}\right\vert \\ 
\text{ \ \ \ \ \ \ \ \ \ \ \ \ \ \ \ \ \ \ \ \ }\leq \frac{19\left(
x+x^{3}\right) y}{36\left( 1+u\right) ^{3}\left( 1+u+r\right) ^{2}}\underset{%
u,r\geq 0}{\sup }\left\{ \left( 1+u+r\right) ^{2}\left\vert w_{2}\right\vert
\right\} \\ 
\text{ \ \ \ \ \ \ \ \ \ \ \ \ \ \ \ \ \ \ \ \ }\leq \frac{19\left(
x^{2}+x^{4}\right) y}{36\left( 1+u\right) ^{3}\left( 1+u+r\right) ^{2}}.%
\end{array}
\tag{4.106}  \label{4.106}
\end{equation}%
Inserting $\left( \ref{4.102}\right) $, $\left( \ref{4.103}\right) $, $%
\left( \ref{4.104}\right) $,\ and$\ \left( \ref{4.106}\right) $ into the\
definition\ of\ $B_{6}$\ in $\left( \ref{4.80}\right) $ we obtain%
\begin{equation}
\begin{array}{l}
\left\vert B_{6}\right\vert \leq \frac{\left( 1+12K_{0}\right) \left(
x^{2p}+x^{2p+2}\right) y}{6\left( 1+u\right) ^{3}\left( 1+u+r\right) ^{2}}+%
\frac{7\left( x^{2}+x^{4}\right) y}{24\left( 1+u\right) ^{3}\left(
1+u+r\right) ^{2}} \\ 
\text{ \ \ \ \ \ \ \ \ \ \ }+\frac{\left( 15+32K_{0}\right) \left(
x^{2}+x^{4}+x^{6}+x^{2p+2}+x^{2p+4}\right) y}{48\left( 1+u\right) ^{3}\left(
1+u+r\right) ^{4}}+\frac{19\left( x^{2}+x^{4}\right) y}{36\left( 1+u\right)
^{3}\left( 1+u+r\right) ^{2}} \\ 
\text{ \ \ \ \ \ \ \ \ \ \ }\leq \frac{\left( 187+384K_{0}\right) \left(
x^{2}+x^{4}+x^{6}+x^{2p}+x^{2p+2}+x^{2p+4}\right) y}{144\left( 1+u\right)
^{3}\left( 1+u+r\right) ^{2}}.%
\end{array}
\tag{4.107}  \label{4.107}
\end{equation}%
Summing up $\left( \ref{4.85}\right) $, $\left( \ref{4.99}\right) $, and $%
\left( \ref{4.107}\right) $,\ considering the\ definition\ of\ $f_{2}$\ in $%
\left( \ref{4.82}\right) $, we arrive at%
\begin{equation}
\begin{array}{l}
\left\vert f_{2}\right\vert =\left\vert B_{4}\overline{\Theta }+B_{5}\left(
w_{2}-\overline{W_{2}}\right) +B_{6}\right\vert \\ 
\text{ \ \ \ \ \ }\leq \frac{\left( 12+56K_{0}\right) \left(
x^{2}+x^{4}+x^{2p}+x^{2p+2}\right) y}{48\left( 1+u\right) ^{3}\left(
1+u+r\right) ^{2}} \\ 
\text{ \ \ \ \ \ }+\frac{\left( 83+160K_{0}\right) \left(
x^{2}+x^{4}+x^{6}+x^{2p+2}+x^{2p+4}\right) y}{12\left( 1+u\right) ^{3}\left(
1+u+r\right) ^{4}} \\ 
\text{ \ \ \ \ \ }+\frac{\left( 187+384K_{0}\right) \left(
x^{2}+x^{4}+x^{6}+x^{2p}+x^{2p+2}+x^{2p+4}\right) y}{144\left( 1+u\right)
^{3}\left( 1+u+r\right) ^{2}} \\ 
\text{ \ \ \ \ \ }\leq \frac{\left( 1219+2472K_{0}\right) \left(
x^{2}+x^{4}+x^{6}+x^{2p}+x^{2p+2}+x^{2p+4}\right) y}{144\left( 1+u\right)
^{3}\left( 1+u+r\right) ^{2}}.%
\end{array}
\tag{4.108}  \label{4.108}
\end{equation}%
Thus, using the same tools as in $\left( \ref{4.30}\right) $, we deduce that 
\begin{equation}
\begin{array}{l}
\int_{0}^{u_{1}}\left\vert \left[ f_{2}\right] _{\gamma _{1}}\right\vert
du=\int_{0}^{u_{1}}\left\vert f_{2}\left( u_{1},r_{1}\right) \right\vert du
\\ 
\text{ \ \ \ \ \ \ \ \ \ \ \ \ \ \ \ \ \ \ \ \ \ \ }\leq \int_{0}^{u_{1}}%
\frac{\left( 1219+2472K_{0}\right) \left(
x^{2}+x^{4}+x^{6}+x^{2p}+x^{2p+2}+x^{2p+4}\right) y}{144\left(
1+u_{1}\right) ^{3}\left( 1+u_{1}+r_{1}\right) ^{2}}du \\ 
\text{ \ \ \ \ \ \ \ \ \ \ \ \ \ \ \ \ \ \ \ \ \ \ }\leq \frac{\left(
1219+2472K_{0}\right) \left(
x^{2}+x^{4}+x^{6}+x^{2p}+x^{2p+2}+x^{2p+4}\right) y}{144}\int_{0}^{u_{1}}%
\frac{1}{\left( 1+u\right) ^{3}\left( 1+u+r\right) ^{2}}du \\ 
\text{ \ \ \ \ \ \ \ \ \ \ \ \ \ \ \ \ \ \ \ \ \ \ }\leq \frac{\left(
1219+2472K_{0}\right) \left(
x^{2}+x^{4}+x^{6}+x^{2p}+x^{2p+2}+x^{2p+4}\right) y}{72\left(
1+u_{1}+r_{1}\right) ^{2}l^{2}\left( x\right) }\int_{0}^{\infty }\frac{1}{%
\left( 1+u\right) ^{3}}du \\ 
\text{ \ \ \ \ \ \ \ \ \ \ \ \ \ \ \ \ \ \ \ \ \ \ }=\frac{\left(
1219+2472K_{0}\right) \left(
x^{2}+x^{4}+x^{6}+x^{2p}+x^{2p+2}+x^{2p+4}\right) y}{144\left(
1+u_{1}+r_{1}\right) ^{2}l^{2}\left( x\right) }.%
\end{array}
\tag{4.109}  \label{4.109}
\end{equation}%
Insertion of $\left( \ref{4.84}\right) $ and $\left( \ref{4.109}\right) $
into $\left( \ref{4.81}\right) $ yields%
\begin{equation}
\begin{array}{l}
\left\vert \vartheta \left( u_{1},r_{1}\right) \right\vert \leq
\int_{0}^{u_{1}}\left\{ \exp \left( \int_{0}^{u_{1}}\left\vert \left[
N_{2}\left( u,r\right) \right] _{\gamma _{1}}\right\vert dv\right) \right\}
\left\vert \left[ f_{2}\right] _{\gamma _{1}}\right\vert du \\ 
\text{ \ \ \ \ \ \ \ \ \ \ \ \ \ \ \ }\leq \frac{\left(
1219+2472K_{0}\right) \left(
x^{2}+x^{4}+x^{6}+x^{2p}+x^{2p+2}+x^{2p+4}\right) y}{144\left(
1+u_{1}+r_{1}\right) ^{2}l^{2}\left( x\right) }\exp \left( K_{1}\left(
x^{2}+x^{4}+x^{2p+2}\right) \right) .%
\end{array}
\tag{4.110}  \label{4.110}
\end{equation}%
Hence%
\begin{equation}
\left\Vert \vartheta \right\Vert _{\mathcal{Y}}\leq \Xi \left( x\right) y, 
\tag{4.111}  \label{4.111}
\end{equation}%
where%
\begin{equation}
\Xi \left( x\right) =\frac{\left( 1219+2472K_{0}\right) \left(
x^{2}+x^{4}+x^{6}+x^{2p}+x^{2p+2}+x^{2p+4}\right) \exp \left( K_{1}\left(
x^{2}+x^{4}+x^{2p+2}\right) \right) }{144l^{2}\left( x\right) }.  \tag{4.112}
\label{4.112}
\end{equation}%
It is easy to see that the function $\Xi $ given\ in\ $\left( \ref{4.112}%
\right) $\ is strictly monotonically increasing on $[0,x_{1}]$ and $\Xi
\left( 0\right) =0$. This shows that there exists $x_{2}\in (0,x_{1}]$ such
that $\Xi \left( x\right) <\frac{1}{2}$ for all $x\in (0,x_{2}]$. Thus, in\
view\ of\ $\left( \ref{4.111}\right) $, the mapping $W\longrightarrow 
\mathcal{K}\left( W\right) $ contracts in $\mathcal{Y}$ for $\left\Vert
W\right\Vert _{\mathcal{X}}\leq x_{2}$. This concludes the proof of the
global existence and uniqueness of classical solution of $\left( \ref{3.8}%
\right) $.

The decay property $\left( \ref{4.4}\right) $ of the solution is a direct
consequence of the definition $\left( \ref{4.1}\right) $ of the Banach
spaces $\left( \mathcal{X},\left\Vert .\right\Vert _{\mathcal{X}}\right) $
and $\left( \mathcal{Y},\left\Vert .\right\Vert _{\mathcal{Y}}\right) $.

Now, from $\left( \ref{3.3}\right) $ and $\left( \ref{4.14}\right) $ one
deduces easily that, for each $r\geq 0$, $g\longrightarrow 1$ if $%
u\longrightarrow \infty $. In view of $\left( \ref{4.20}\right) $, this
implies that, for each $r\geq 0$, $\widetilde{g}\longrightarrow 1$ if $%
u\longrightarrow \infty $. So, as $u\longrightarrow \infty $, the metric
given in Bondi coordinates by $\left( \ref{2.8}\right) $ becomes the
Minkowski metric. With this, we are done with the proof of Theorem 4.1.

\begin{remark}
$\left( i\right) $ Theorem 4.1 was stated and proved, under the more
restrictive assumption $p\geq 3$, by\ Chae \cite{4} for the
Einstein-Maxwell-Higgs system. Note that the solution obtained here decays
more\ slowly than that of \cite{6} concerning the spherically symmetric
massless Einstein-scalar field system. We found out that this latter fact
stems essentially from the estimate (see \ref{4.17}) 
\begin{equation*}
\left\vert \frac{Q}{r}\right\vert \leq \frac{x^{2}}{2\left( 1+u\right)
^{2}\left( 1+u+r\right) },
\end{equation*}%
due to the non-vanishing of the local charge $Q$. Some questions raised in 
\cite{4} are thus answered. It would be interesting to find out whether one
can use conformal compactification methods of Penrose \cite{23} to explain
slow decaying of the solution via extension by continuity to conformal null
infinity.

$\left( ii\right) $ Theorem 4.1 easily applies so as to encompass global
existence and uniqueness of classical solutions of the spherically symmetric
EYMH system with vanishing self-interaction potential $V$ and those of the
non linear Einstein-Klein-Gordon system as well. Moreover, in the latter
case, it turns out that the solutions possess the same order of decay
estimates as those obtained in \cite{6}.

$\left( iii\right) $ It is worth mentioning that assumption $\left( \ref{4.2}%
\right) $ is not fulfilled by the (classical) self-interaction potential $%
V\left( t\right) =t^{2}$.
\end{remark}

\textbf{Acknowledgements. }C Tadmon wishes to express sincere thankfulness
to Professor Mamadou Sango for warm welcome at the University of Pretoria
where this work was finalized at the beginning of his postdoctoral
fellowship.

\end{document}